\documentclass[final,3p,times]{elsarticle}
\usepackage{graphicx}
\usepackage{amssymb}
\usepackage{amsmath}
\usepackage{mathtools}
\usepackage{bm}
\usepackage[hyphens]{url}
\usepackage{hyperref}
\usepackage{algorithm}
\usepackage{algpseudocode}
\hypersetup{pdfauthor=author}
\usepackage{lineno}
\usepackage{booktabs}
\usepackage{adjustbox}
\usepackage{subcaption}
\usepackage{lipsum}
\usepackage{multirow}
\usepackage{xcolor}
\usepackage{algorithm}
\DeclareMathOperator*{\argmin}{argmin}

\journal{Journal of Computational Physics}

\begin{document}

\begin{frontmatter}


\title{Data-driven reduced-order modelling for blood flow simulations with geometry-informed snapshots}



\author[utaddress,uvaaddress]{Dongwei Ye\corref{mycorrespondingauthor}}
\cortext[mycorrespondingauthor]{Corresponding authors}
\ead{d.ye-1@utwente.nl}

\author[uvaaddress]{Valeria Krzhizhanovskaya}
\ead{v.krzhizhanovskaya@uva.nl}

\author[uvaaddress]{Alfons G. Hoekstra}
\ead{a.g.hoekstra@uva.nl}

\address[utaddress]{Mathematics of Imaging \& AI, Department of Applied Mathematics, University of Twente, The Netherlands}
\address[uvaaddress]{Computational Science Lab, Informatics Institute, Faculty of Science, University of Amsterdam, The Netherlands}

\begin{abstract}
Parametric reduced-order modelling often serves as a surrogate method for hemodynamics simulations to improve the computational efficiency in many-query scenarios or to perform real-time simulations. However, the snapshots of the method require to be collected from the same discretisation, which is a straightforward process for physical parameters, but becomes challenging for geometrical problems, especially for those domains featuring unparameterised and unique shapes, e.g. patient-specific geometries. In this work, a data-driven surrogate model is proposed for the efficient prediction of blood flow simulations on similar but distinct domains. The proposed surrogate model leverages group surface registration to parameterise those shapes and formulates corresponding hemodynamics information into geometry-informed snapshots by the diffeomorphisms constructed between a reference domain and original domains. A non-intrusive reduced-order model for geometrical parameters is subsequently constructed using proper orthogonal decomposition, and a radial basis function interpolator is trained for predicting the reduced coefficients of the reduced-order model based on compressed geometrical parameters of the shape. Two examples of blood flowing through a stenosis and a bifurcation are presented and analysed. The proposed surrogate model demonstrates its accuracy and efficiency in hemodynamics prediction and shows its potential application toward real-time simulation or uncertainty quantification for complex patient-specific scenarios. 
\end{abstract}


\end{frontmatter}

% \linenumbers
% \numberwithin{table}{section}
% \numberwithin{figure}{section}

\section{Introduction}
The mechanical and biological interaction between blood flow and the vessel wall has a significant impact on the initiation and progression of vascular diseases, e.g. the development of atherosclerotic plaque \cite{Maurits1998,Saurabh2010,Habib2011}, or restenosis of arteries after percutaneous intervention \cite{Csaba2016,Konstantinos2012}. In-silico simulations have demonstrated significant efficiency and flexibility in those studies of hemodynamics, offering deep insights into mechanisms of disease, biological/pathological processes, and designs of medical devices \cite{OWIDA2012689,ISR3D2017,Gabor2017,Corti2021}. Therefore computational fluid dynamics techniques are widely applied to perform high-fidelity simulations to mimic and predict the behavior of blood flow, to further support the clinical practice or in-depth study of physiology and pathology \cite{Martin2009,ZUN2021110361,Zhang2014,Dur2011}.

However, owing to the complexity and scale of biofluid problems, the evaluation of the model using numerical methods, such as finite element methods (FEM), could be computationally expensive or even becomes prohibitive in the many-query scenarios, such as design and optimisation, reliability analysis and uncertainty quantification \cite{Ye2022,ArzaniAmirhossein2021,YE2021107734}. Reduced order modelling (ROM) provides a high-fidelity framework to reduce the computational complexity for solving parametric partial differential equations (PDEs) \cite{ArzaniAmirhossein2021,Jian2019}. It exploits the intrinsic correlation of the full-order model (FOM) solution over a physical domain, time evolution, or parameter space, and projects the system to a low-dimensional reduced space. Owing to efficient and reliable prediction based on the reduced systems, such surrogate models also enable the real-time performance of the simulation \cite{Niroomandi2012,Aguado2015,Fresca2022}. One of the widely applied reduced-order methods is Proper Orthogonal Decomposition (POD) \cite{POD2003}. The POD can be realised using principal component analysis (PCA), or the singular value decomposition (SVD). The projection coefficients can be computed either in an intrusive manner by manipulating the governing equations with Galerkin methods \cite{ROWLEY2004115,WANG2020109402} or in a non-intrusive manner by formulating it as an interpolation problem \cite{HESTHAVEN201855,Jian2019}. Dynamic mode decomposition is another projection-based method for the temporal decomposition of fluid dynamics \cite{schmid2010,Proctor2016,Kutz2016}. The empirical interpolation method \cite{BARRAULT2004667} and the discrete empirical interpolation method \cite{Chaturantabut2010} were proposed to recover an affine expansion for the nonlinear problem. 

One of the essential parts of parametric ROM is collecting snapshots from high-fidelity simulations. The magnitudes of these snapshots reflect the influence of parameter variations on the solution manifolds. The collecting procedure is straightforward for the problems related to physical parameters since the snapshots are generated from the simulations sharing the same discretisation. However, it becomes challenging for geometrical problems, especially those associated with domains featuring unparameterised and unique shapes, a common scenario in hemodynamic simulations. The geometric shapes of domains in hemodynamic simulations are either generated artificially based on a simplification of the anatomical geometry or directly segmented and reconstructed from clinical image data \cite{quarteroni2019mathematical}, i.e. patient-specific geometries. These anatomical geometries of a kind, such as the segments of arteries and organs are similar in terms of general shape but different in details. In order to leverage those snapshots associated with different discretisations to construct parametric ROM for flow field predictions under geometric variation, two issues must be addressed. First, the spatial compatibility over the snapshots has to be guaranteed for the construction of reduced bases. Meanwhile, the geometric shapes of domains have to be parameterised in a consistent manner such that the mapping between the geometric parameters and reduced coefficients of flow bases can be approximated, and subsequently utilised in the prediction.

The existing methods primarily seek a corresponding transformation between distinct geometrical configurations, i.e. original domains and a reference domain, via shape reconstruction, and embed the transformation into the parametric PDEs such that all the snapshots are generated on a single reference domain. Parameter identification methods, e.g. radial basis functions interpolation \cite{Manzoni2012,Manzoni2014}, inverse distance weighting interpolation \cite{Ballarin2019} or free-form deformation \cite{Manzoni2012Shapeoptimzation,RozzaHessStabile2021,LASSILA20101583} are applied to parameterise the geometries before the shape reconstruction. However, those parametrisation methods heavily depend on the selection of control points, the smoothness of polynomials and the choice of the reference shape. The complexity of the parametrisation increases significantly when dealing with intricate geometries, such as patient-specific data, and consequently influences the precision of shape reconstruction.  

In this work, a data-driven surrogate model based on the non-intrusive reduced order modelling method and surface registration is proposed. The surrogate model aims to approximate the fluid dynamics of blood flow within domains of distinct but similar shapes efficiently. Different from the existing approaches mentioned above, the proposed surrogate model performs surface registrations based on the form of currents to approximate the diffeomorphism between the shapes of the original domains and the reference shape. The currents-based registration method enables registration without point-to-point correspondence \cite{Vaillant2005,DURRLEMAN2009793} and does not require upfront parametrisation. On the contrary, the registration process itself offers a means to parameterise these shapes, obviating additional procedures. The diffeomorphisms constructed during surface registrations provide the mappings between the reference domain and the original domains. The evaluations are subsequently performed on the reference domain by projection using the diffeomorphisms to ensure the spatial compatibility over the snapshots, namely geometry-informed snapshots. Therefore a reduced-order model of geometrical parameters can be formulated. POD is applied to construct reduced bases and the reduced coefficients are computed via a radial basis function (RBF) interpolation method. A schematic diagram of the proposed surrogate model is shown in Figure~\ref{fig:ROMSR}. The proposed data-driven reduced-order model with geometry-informed snapshots resolves the spatial compatibility and parametrisation issues for geometric ROM problems and enables efficient and accurate prediction of hemodynamics within similar geometries of domains.

% The paper is arranged as follows. The details of the surrogate model are described in Section~\ref{sec:Method}. Two numerical examples are presented and discussed in Section~\ref{sec:example} and Section~\ref{sec:discussion}, followed by a conclusion in Section~\ref{sec:conclusion}.

\section{Method}\label{sec:Method}
\subsection{Surface registration with currents}
Here we introduce an approach based on non-parametric representation of surfaces using currents and its deformation framework \cite{Vaillant2005,DURRLEMAN2009793,Durrleman2010}. The method characterises shapes in the form of currents which enables quantification of the dissimilarity between two shapes without point correspondences, and subsequently can be applied in registration.  We assume that the shapes of domains in the blood flow simulations are similar and can be achieved by deformation from a reference domain. A diffeomorphic registration, which transforms shapes from a source coordinate system to a target coordinate system, is performed via minimising the difference between two surfaces and geodesic energy.

\subsubsection{Represent surfaces using currents}
Given a continuous vector field $\bm{\omega}: \mathbb{R}^d \rightarrow \mathbb{R}^d$ in the ambient space $\mathbb{R}^d$, where $d=2 \text{ or } 3$, an oriented piecewise-smooth surface $S$ in $\mathbb{R}^3$ can be characterised by the flux going through the surface,
\begin{equation}\label{eq:chap4_flux}
    [S](\bm{\omega}) = \int_{S} \langle \bm{\omega}(\bm{x}), \bm{n}(\bm{x}) \rangle dS(\bm{x}),
\end{equation}
where $\bm{n}(\bm{x})$ denotes the unit normal vector of the surface $S$ at point $\bm{x}$. For a curve in the two-dimensional case ($d=2$),  $\bm{n}(\bm{x})$ is replaced by a tangent vector. The computation of the flux defines a linear mapping from a space of vector field $W$ to $\mathbb{R}$. This mapping is called the current associated with surface $S$. We denote the space of currents as $W'$. By defining the space $W$ as a Reproducing Kernel Hilbert Space (RKHS), the vector field can be represented in the form of a kernel function $K:\mathbb{R}^3 \times \mathbb{R}^3 \rightarrow \mathbb{R}$, therefore a vector field $\bm{\omega}$ in $W$ can be reformulated with fixed points $\{\bm{x}^*_i\}_{i=1,2,\cdots,n}$ and their corresponding vectors $\beta$,
\begin{equation}\label{eq:chap4_kernel}
    \bm{\omega}(\bm{x}) = K^W(\bm{x},\bm{x}^*)\beta.
\end{equation}
The space $W$ equips with inner product $\langle K(\cdot,\bm{x})\alpha,K(\cdot,\bm{x}')\beta\rangle_{W} = \alpha^\text{T}K(\bm{x},\bm{x}')\beta$. Generally a Gaussian kernel $ K^{W}(x_i,x_j)=\text{exp}(-\frac{|x_i-x_j|^2}{\lambda_W})$ is applied, where $\lambda_W$ denotes the scale at which $W$ may spatially vary. By substituting the expression of equation~\ref{eq:chap4_kernel}, the inner product of space $W$ can be written as $\langle K(\cdot,\bm{x})\alpha,\bm{\omega}\rangle_{W} = \alpha^\text{T}\bm{\omega}(\bm{x})$, which is known as reproducing property. Consequently, equation~\ref{eq:chap4_flux} can be reformulated into,
\vspace{-0.3cm}
\begin{equation}\label{eq:chap4_kernelflux}
\arraycolsep=1.4pt\def\arraystretch{2}
\begin{array}{rll}
    [S](\bm{\omega})& = \int_{S} \langle \bm{\omega}, K(\cdot,\bm{x})\bm{n}(\bm{x}) \rangle_W dS(\bm{x})\\
    &= \left\langle \int_{S} K(\cdot,\bm{x})\bm{n}(\bm{x})dS(\bm{x}), \bm{\omega}  \right\rangle_W. 
\end{array}
\end{equation}
It shows that $\int_{S} K(\cdot,\bm{x})\bm{n}(\bm{x})dS(\bm{x})$ is a unique Riesz representation in $W$ of current $[S]$ in $W'$ \cite{CHARON2020441}. Therefore, the inner produce of two currents $[S_1]$ and $[S_2]$ can be obtained by,
\begin{equation}
    \langle [S_1],[S_2] \rangle_{W'} = \int_{S_1}\int_{S_2} K(\bm{x},\bm{x}')\langle \bm{n}_1(\bm{x}),\bm{n}_2(\bm{x}') \rangle
    dS_{1}(\bm{x})dS_2(\bm{x}').
\end{equation}
The associated norm can be applied to measure the dissimilarity between two shapes,
\begin{equation}
   \lVert [S_1] -[S_2] \rVert^2_{W'} =  \langle [S_1],[S_1] \rangle^2_{W'} - 2 \langle [S_1],[S_2] \rangle_{W'} + \langle [S_2],[S_2] \rangle^2_{W'} .
\end{equation}

\subsubsection{Diffeomorphism construction}
The shape of a reference domain can be represented by a series of control points (CPs) $\{\bm{x}_i^{\text{cp}}\}_{i=1,...,N_{\text{cp}}}$ on a boundary (e.g. the boundary vertices in a finite element mesh). The surface registration of a reference shape $S_{\text{ref}}$ to a target shape $S_{\text{tar}}$ can be performed by finding a flow of diffeomorphism $\varphi(\cdot,t): \mathbb{R}^d \rightarrow \mathbb{R}^d$ such that the dissimilarity of the shapes, as well as geodesic energy of the process, are minimised. The trajectory of these points at a given artificial time $t$, can be described using a flow differential equation:
\begin{equation}\label{eq:diffvelo}
    \frac{\partial\varphi(\bm{x},t)}{\partial t} = \bm{v}(\varphi(\bm{x},t)),
\end{equation}
where $\bm{v}(\varphi(\bm{x},t))$ is a vector field representing the velocity of the deformation. Similar to equation~\ref{eq:chap4_kernel}, the velocity vector field can be approximated by a summation of kernel function at discrete CPs,
\begin{equation}
\bm{v}(\bm{x}) = \sum_{i=1}^{N_{\text{cp}}} K(\bm{x},\bm{x}^{\text{cp}}_i)\bm{\alpha}_i,    
\end{equation}
where $\bm{\alpha}_i$ denotes the basis vector of the field and $K(\cdot,\cdot)$ is a kernel function that decides the weights of each basis based on the distance between two points. Together with equation~\ref{eq:diffvelo}, the flow of diffeomorphism can be reformulated as the motion of CPs,
\begin{equation}\label{eq:flowofdiffeo}
    \frac{\partial\varphi_t(\bm{x}_j)}{\partial t} = \sum_{i=1}^{N_{\text{cp}}} K(\bm{x}_j,\bm{x}^{\text{cp}}_i)\bm{\alpha}_i(t), \; j = 1,\cdots,N_{\text{cp}}.
\end{equation}
To perform a surface registration from a reference shape $S_{\text{ref}}$ to a target shape $S_{\text{tar}}$, a proper velocity vector field can be computed via an optimisation problem formulated as,
\begin{equation}\label{eq:velocityoptimzation}
    \argmin_{\bm{v}} \textrm{J}(\bm{v}) = \lVert [\varphi(S_{\text{ref}})] - [S_{\text{tar}}] \rVert^2_{W'} + 
    \int_0^1 \lVert\bm{v} \rVert^2_W \textrm{d}t,
\end{equation}
where $ \lVert [\varphi(S_{\text{ref}})] - [S_{\text{tar}}] \rVert^2_{W'}$ measures the difference between the deformed shape and target shape using currents; $\int_0^1 \lVert\bm{v} \rVert^2_W \textrm{d}t$ denotes the total kinetic energy required to deform the shape from its initial state. For further details of surface registration with currents, see \cite{Durrleman2010,CHARON2020441}.

Note that the direct outcome of equation~\ref{eq:flowofdiffeo} is the position of the CPs after the diffeomorphism $\varphi(\bm{x}_i^{\text{cp}})$ rather than the complete diffeomorphism itself. Therefore, the diffeomorphism has to be approximated by leveraging the known data $\{(\bm{x}_i^{\text{cp}},\varphi(\bm{x}_i^{\text{cp}}))\}_{i=1}^{N_{\text{cp}}}$. Such an approximation is achieved by radial basis function interpolation. We denote the approximation function as $\mathcal{X}(\cdot;\bm{\gamma})$, where $\bm{\gamma}$ is the associated geometric parameters discussed in the next section.

\subsection{Parametrisation of geometries}
The constructed diffeomorphism has a twofold application in the proposed method. First, if the domains of different shapes can be achieved by diffeomorphisms from the same reference domain, those shapes can therefore be parameterised by the displacement of the CPs of the reference shape before and after the deformation. Consider the shape of the reference domain is represented by a set of CPs, $X_* = \{\bm{x}_{*,i}\}_{i=1}^{b_*} \in \mathbb{R}^{b_* \times d} $, where $b_*$ and $d$ denote the number of CPs and spatial dimension ($d=2$ or 3), respectively. For any target shape considered as a result of deformation of the reference shape by a particular diffeomorphism $\varphi$, the geometrical parametrisation of such a shape can be expressed as,
\begin{equation}
    \bm{\gamma} = [\bm{e}_1^{\mathrm{T}}(\varphi(X_*) - X_*)^{\mathrm{T}},\cdots, \bm{e}_d^{\mathrm{T}}(\varphi(X_*) - X_*)^{\mathrm{T}}]^{\mathrm{T}}
\end{equation}
where $\bm{\gamma} \in \mathbb{R}^{db_*}$ and $\{\bm{e}_i\}_{i=1}^d$ denote unit vectors of $d$-dimensional Euclidean space. The expression essentially means the geometry is parameterised by a stack of displacements of the CPs (of the reference shape) in each spatial dimension. However, such an approach to parameterise the shape may lead to the high dimensionality of parameter space owing to the number of the CPs of a two- or three-dimensional shape. Besides, correlations can be found between the displacements of those CPs. Therefore, a dimensionality reduction technique can be employed to further extract the latent low-dimensional representation of these parameters,
\begin{equation}\label{eq:reducedgeoparameter}
    \bm{\gamma} \approx \sum_{i=1}^{N_k} \gamma^{\text{rb}}_i \bm{\varrho}_i = Q \bm{\gamma}^{\mathrm{rb}},
\end{equation}
where $N_k$ denotes the number of the truncated reduced basis of geometrical parameters. $\gamma_i^{\mathrm{rb}}$ and $\bm{\varrho}_i$ denote the reduced coefficients and reduced bases of the geometrical parameters, respectively. $Q = [\bm{\varrho_1}|\cdots|\bm{\varrho}_{N_k}] \in \mathbb{R}^{b_*\times N_k}$ and $\bm{\gamma}^{\mathrm{rb}} = [\gamma^{\mathrm{rb}}_1,\cdots,\gamma^{\mathrm{rb}}_{N_k}]^{\mathrm{T}} \in \mathbb{R}^{N_k}$ are corresponding matrix and vector representation. POD is applied to extract the reduced basis of geometrical parameters. Consequently, the reduced coefficients of the geometrical parameters $\bm{\gamma}^{\text{rb}}$ can be viewed as a compressed representation of $\bm{\gamma}$ and $N_k \ll db_*$. In the construction of non-intrusive ROM for blood flow simulation, we consider the reduced coefficients of the flow fields as a latent function of reduced geometric parameters $\bm{\gamma}^{\text{rb}}$ and train an RBF interpolator for the corresponding prediction.

\subsection{Geometry-informed snapshots based on FOM on a reference domain}\label{sec:chap4_FOM}
Apart from parametrisation, the constructed diffeomorphisms are also applied to generate all the geometry-informed snapshots with different shapes of domains by projecting and solving the FOMs onto a reference domain. Consider blood flow with a moderate Reynolds number as an incompressible Newtonian fluid to be modelled with the steady Navier-Stokes equations. Within a parametric vessel lumen as the spatial domain $\Omega(\bm{\gamma}) \in \mathbb{R}^d$, where $d=2$ or $3$, such problem can be formulated as: find the vectorial velocity field $\bm{u}(\bm{x};\bm{\gamma}): \Omega(\bm{\gamma}) \rightarrow \mathbb{R}^d$, and scalar pressure field $p(\bm{x};\bm{\gamma}): \Omega(\bm{\gamma}) \rightarrow \mathbb{R}$, such that:

\begin{equation}\label{eq1}
\begin{array}{rcl}
    (\bm{u} \cdot \nabla) \bm{u} - \nu \nabla^2\bm{u} + \nabla p &=\bm{\mathrm{f}}, \\
    
    \nabla \cdot \bm{u} &= 0,
\end{array}    
\end{equation}
and subject to boundary conditions:
\begin{subequations}
\begin{align}
     \bm{u}  & = \bm{\mathrm{g}}_{\text{D}}, \hspace{2cm} \bm{x} \in \partial\Omega_{\text{D}}(\bm{\gamma}) \label{eq2:sub1},\\ 
     -p\bm{n}+ \nu (\bm{n}\cdot \nabla) \bm{u} & = \bm{\mathrm{g}}_{\text{N}}, \hspace{2cm} \bm{x} \in \partial\Omega_{\text{N}}(\bm{\gamma}) \label{eq2:sub2}, 
\end{align}
\end{subequations}
where $\bm{\gamma} \in \mathcal{P} \subset \mathbb{R}^{N_p}$ stands for the geometric parameters; $\nu$ denotes the kinematic viscosity; $\bm{\mathrm{f}}$ denotes the body force. Equation~\ref{eq2:sub1} and \ref{eq2:sub2} represent Dirichlet and Neumann boundary condition on $\partial\Omega_{\text{D}}$ and $\partial\Omega_{\text{N}}$, respectively. Dirichlet boundary condition is generally defined for the inlet and wall boundary of a vessel, while the Neumann boundary condition is applied to the outlet. For the sake of simplicity, the body force term is taken to be zero. 

As mentioned before, the governing equations should be solved on a parameter-independent reference domain $\Omega^*$ to ensure the spatial compatibility of the snapshot through a parameterised mapping $\mathcal{X}(\bm{\xi} ; \bm{\gamma})$, i.e. $\bm{x} = \mathcal{X}(\bm{\xi};\bm{\gamma})$. 
The variational form of the system is therefore obtained by introducing two functional spaces $\mathcal{V} \coloneqq \{ \bm{w} \in \mathcal{H}^1(\Omega^*) \mid \bm{w} = 0 \; \text{on} \; \partial\Omega^*_{\text{D}})\}$ of weighting functions for momentum conservation and $\mathcal{Q} \coloneqq \{ q \in \mathcal{L}^2(\Omega^*)\}$ for incompressible constraint over the reference domain. $\mathcal{Q}$ can be also applied as the solution space for pressure. 
Besides, solution spaces $\mathcal{S} \coloneqq \{ \bm{u} \in \mathcal{H}^1(\Omega^*) \mid \bm{u} = \bm{g}_{\text{D}}  \; \text{on} \; \partial\Omega^*_{\text{D}} \}$ 
% and $\mathcal{T} \coloneqq \{ p \in \mathcal{L}^2(\Omega(\bm{\gamma})) \}$
is defined over the domain for velocity. The governing equations are then projected to the weighting spaces by multiplying weighting functions and integrating over the domain, which reformulates the problem \ref{eq1} to: find $\bm{u}(\bm{\xi}) \in \mathcal{S}$ and $p(\bm{\xi}) \in \mathcal{Q}$ such that for all $\bm{w}$ and $q$,
\begin{equation}\label{eq:snapshotgeneration}
\begin{array}{rll}
    a(\bm{w},\bm{u};\bm{\gamma}) + c(\bm{u},\bm{w},\bm{u};\bm{\gamma}) + b(\bm{w},p;\bm{\gamma}) =& (\bm{w},\bm{\mathrm{g}}_{\text{N}}), \\
    
    b(\bm{u},q;\bm{\gamma}) =& 0
\end{array}    
\end{equation}
where 
% \vspace{-0.05cm}
\begin{equation*}
a(\bm{w},\bm{u};\bm{\gamma}) = \int_{\Omega^*} \sum_{i=1}^d \sum_{j=1}^d \frac{\partial\bm{u}}{\partial \xi_i}
\kappa_{ij}(\bm{\xi};\bm{\gamma})
\frac{\partial\bm{w}}{\partial \xi_j} \mathrm{d}\Omega^*,
 \hspace{1cm}
b(\bm{w},p;\bm{\gamma}) = -\int_{\Omega^*}p \sum_{i=1}^d \sum_{j=1}^d 
\zeta_{ij}(\bm{\xi};\bm{\gamma})
\frac{\partial w_j}{\partial \xi_i} 
\mathrm{d}\Omega^*,
\end{equation*}

% \begin{equation*}
% \textcolor{blue}{b(\bm{u},q;\bm{\gamma}) = -\int_{\Omega^*}q \sum_{i=1}^d \sum_{j=1}^d 
% \zeta_{ij}(\bm{\xi};\bm{\gamma})
% \frac{\partial u_j}{\partial \xi_i} 
% \mathrm{d}\Omega^*,},
% \end{equation*}
\begin{equation*}
c(\bm{u},\bm{w},\bm{u};\bm{\gamma}) =  \int_{\Omega^*} \sum_{i=1}^d \sum_{j=1}^d \sum_{k=1}^d 
u_i \zeta_{ij}(\bm{\xi};\bm{\gamma})
\frac{\partial w_k}{\partial \xi_j}u_k
\mathrm{d}\Omega^*.
\end{equation*}
These are the bilinear and trilinear forms for diffusion term, pressure-divergence term and convection term of the system, respectively. The elements $\kappa_{ij}(\bm{\xi};\bm{\gamma})$ and $\zeta_{ij}(\bm{\xi};\bm{\gamma})$ in the compact forms come from the tensor representation of functions associated with coordinate transformation,
\begin{subequations}
\begin{align}
    \bm{\kappa}(\bm{\xi},\bm{\gamma}) &= \nu (J_{\mathcal{X}}(\bm{\xi};\bm{\gamma}))^{-1} (J_{\mathcal{X}}(\bm{\xi};\bm{\gamma}))^{-\mathrm{T}}
    |J_{\mathcal{X}}|, \\
    \bm{\zeta}(\bm{\xi},\bm{\gamma}) &=(J_{\mathcal{X}}(\bm{\xi};\bm{\gamma}))^{-1}
    |J_{\mathcal{X}}|,
\end{align}
\end{subequations}
where $J_{\mathcal{X}}(\bm{\xi};\bm{\gamma})$ denotes Jacobian matrix of mapping $\mathcal{X}(\cdot;\bm{\gamma})$ and $|J_{\mathcal{X}}|$ is the determinant of the matrix.

Galerkin spatial discretisation approximates the solution by seeking solutions within corresponding finite dimensional subspace $\mathcal{V}^{h} \subset \mathcal{V}$ and $\mathcal{Q}^h \subset\mathcal{Q}$, and the problem is subsequently reformulated to: seek velocity field $\bm{u}^h \in \mathcal{S}^h$ and pressure $p^h \in \mathcal{Q}^h$, such that for all $(\bm{w}^h,q^h) \in \mathcal{V}^h \times \mathcal{Q}^h$,
\begin{equation}
\begin{array}{rll}
    a(\bm{w}^h,\bm{u}^h;\bm{\gamma}) + c(\bm{u}^h,\bm{w}^h,\bm{u}^h;\bm{\gamma}) + b(\bm{w}^h,p^h;\bm{\gamma}) &= (\bm{w}^h,\bm{\mathrm{g}}^h_{\text{N}}), \\
    
    b(\bm{u}^h,q^h;\bm{\gamma}) &= 0.
\end{array}    
\end{equation}
The discrete solution of finite element approximation to the problem can be obtained by solving the system above with the given boundary conditions. The computational accuracy of the discretised system derived is related to the degrees of polynomial chosen for shape functions and the size of the elements. In the remaining part of the paper, we denote the discrete solutions in general of a FOM blood flow simulation with geometric parameters $\bm{\gamma}$ on a reference domain $\Omega^*$ as $\mathbf{u}^h(\bm{\xi};\bm{\gamma})$, namely geometry-informed snapshots in context of ROM. 

\subsection{Reduced order model}
ROM seeks hidden low-dimensional patterns behind the FOM and approximates the FOM with high fidelity. Here ROM is applied to extract the low-dimensional representation over geometric parameter space and the FOM solution therefore can be approximated by,
\begin{equation}\label{eq:ROM}
   \mathbf{u}^h(\bm{\xi};\bm{\gamma}) \approx \hat{\mathbf{u}}^h(\bm{\xi},\bm{\gamma}) = \sum_{i=1}^{N_l} \text{u}^{\text{rb}}_{i}(\bm{\gamma}) \bm{\phi}_i(\bm{\xi}) = \bm{\Phi}\bm{\mathrm{u}}^{\mathrm{rb}}(\bm{\gamma}),
\end{equation}
where $\bm{\mathrm{u}}^{\mathrm{rb}}(\bm{\gamma}) = [\text{u}^{\text{rb}}_{1}(\bm{\gamma}),\cdots,\text{u}^{\text{rb}}_{N_l}(\bm{\gamma})]^{\mathrm{T}}$ denotes the reduced coefficients which provides the weighting of each reduced basis and $\bm{\Phi} = [\bm{\phi}_1(\bm{\xi})|\cdots|\bm{\phi}_{N_l}(\bm{\xi})]$ is the corresponding reduced basis. The decomposition can be applied to any flow fields such as velocities in different directions, pressure, and shear stress fields. We denote $N$ and $N_{\gamma}$ as the dimension of snapshots and the number of snapshots available (corresponding to geometrical parameters), respectively in the ROM. 

\subsubsection{Proper orthogonal decomposition}\label{sec:POD}
Consider a snapshot matrix $\bm{\mathrm{M}}\in \mathbb{R}^{{N\times N_{\gamma}}}$ consisting of snapshots with respect to corresponding geometric parameters $
    \bm{\mathrm{M}} = [\mathbf{u}^h(\bm{\gamma}_1)|
    \mathbf{u}^h(\bm{\gamma}_2)|\cdots|
    \mathbf{u}^h(\bm{\gamma}_{N_{\gamma}})].$
% We denote the centered snapshot matrix as $\Tilde{\bm{\mathrm{M}}}$. 
By performing SVD, such a snapshot matrix can be written in a factorization form,
\begin{equation}
    \bm{\mathrm{M}} = \textbf{\textrm{Z}} \bm{\Sigma} \textbf{\textrm{V}}^{\textrm{T}},
\end{equation}
where $\textbf{\textrm{Z}}$ and $\textbf{\textrm{V}}$ denote left and right orthonormal matrices. The diagonal matrix $\bm{\Sigma}$ consists of singular values $\sigma_i,  \text{where} \hspace{0.1cm} i = 1,...,{N_{\gamma}}$ and $\sigma_1 \geq \sigma_2\geq ... \geq \sigma_{{N_{\gamma}}} \geq 0$.
The objective of POD is to find out a set of orthogonal basis $\bm{{\Phi}} = \{ \bm{\phi}_1, \bm{\phi}_2, \cdots, \bm{\phi}_{N_l} \}$  from the space $\mathcal{Z} = \{ \Tilde{\bm{\Phi}} \in \mathbb{R}^{N \times N_l} : \Tilde{\bm{\Phi}}^\mathrm{T} \Tilde{\bm{\Phi}} =\textbf{\textrm{I}} \}$  containing all possible orthogonal bases, such that the projection residual is minimised:
\begin{equation}\label{eq:orthbasis}
    \min_{\Tilde{\bm{\Phi}} \in \mathcal{Z}} \lVert \bm{\mathrm{M}} -  \Tilde{\bm{\Phi}} \Tilde{\bm{\Phi}}^{\mathrm{T}} \bm{\mathrm{M}} \rVert ^2 _{F} ,
\end{equation}
where $\lVert \cdot \rVert_{F}$ denotes the Frobenius norm. The Eckart-Young theorem \cite{Eckart1936} shows that the orthogonal bases constructed by the basis vectors $\{\bm{z}_i\}_{i=1}^{N_l}$ taken
from the $i$th column of $\textbf{\textrm{Z}}$ is the unique solution to the optimization problem. The cumulative energy captured by $N_l$ number of reduced bases can be evaluated by the ratio between truncated and complete singular values,
\begin{equation}
   \mathcal{R}_{\text{en}} = \frac{\lVert \bm{\mathrm{M}} - \bm{\Phi} \bm{\Phi}^{\mathrm{T}} \bm{\mathrm{M}} \rVert ^2 _F  }{\lVert \bm{\mathrm{M}} \rVert ^2 _{F}} = \frac{\sum^{N_l}_{i=1} \sigma_i^2}{\sum^{N_{\gamma}}_{i=1} \sigma_i^2}.
\end{equation}
The values of $\sigma_i$ decay rapidly which means a small number of bases are sufficient to approximate the data with sufficient accuracy.   

\subsubsection{Radial basis function interpolation}\label{sec:chap4_rbf}
The RBF interpolation method is one of the state-of-the-art methods for multivariate interpolation \cite{wright2003radial,Morris2008}. The method approximates the desired function by a linear combination of basis functions based on known collocation points. See \cite{wright2003radial} for more technical details.

In the framework of non-intrusive ROM, RBF interpolation is applied for approximating the mapping between the reduced geometric parameters $\bm{\gamma}^{\mathrm{rb}}$ and the reduced coefficients of flow fields $\bm{\mathrm{u}}^{\mathrm{rb}}$. Consider $f: \mathbb{R}^{N_k} \rightarrow \mathbb{R}^{N_l}$ to be the latent functions between $\bm{\gamma}^{\mathrm{rb}}$ and $\bm{\mathrm{u}}^{\mathrm{rb}}$. The RBF interpolator $\hat{f}$ can be trained by a collection of training samples $\{(\bm{\gamma}^{\mathrm{rb}}_i,\bm{\mathrm{u}}^{\mathrm{rb}}(\bm{\gamma}_i^{\mathrm{rb}})) \}_{i=1}^{\ell}$,
\begin{equation}\label{eq:chap4_RBFI}
    \hat{f}(\bm{\gamma}^{\mathrm{rb}}) = \sum_{i=1}^{\ell} \lambda_i \psi(\| \bm{\gamma}^{\mathrm{rb}} -\bm{\gamma}^{\mathrm{rb}}_i \|) + P(\bm{\gamma}^{\mathrm{rb}}),
\end{equation}
such that 
$\hat{f}(\bm{\gamma}^{\mathrm{rb}}_j) =  \bm{\mathrm{u}}^{\mathrm{rb}}_j, \; \text{for } j=1,\cdots,\ell,$
where $\psi:[0,+\infty] \rightarrow \mathbb{R}$ is the radial basis function; $\lambda_i$ denotes the expansion coefficients (weights) for each collocation point; $P(\bm{x})$ denotes a low-order polynomial function for conditionally positive definite radial functions.

\subsubsection{Online-offline framework of ROM with geometry-informed snapshots}

\begin{algorithm}[b!]
% \setstretch{1.15}
\caption{Offline stage of the surrogate model}
\label{alg:PODoffline}

\textbf{Input}: Shapes of training domains $\{\bm{X}_j\}_{j=1}^{N_s} = \{\{\bm{x}_{1,i}\}^{b_{1}}_{i=1},\cdots,\{\bm{x}_{{N_s},i}\}^{b_{N_s}}_{i=1} \}$; Shape of the reference domain $\bm{X}_*=\{\bm{x}_{*,i}\}^{b_{*}}_{i=1}$ \\
\textbf{Output}: Reduced bases of geometric parameters $Q$; Reduced bases of flow fields $\bm{\Phi}$; RBF-ROM interpolator, $\hat{f}(\bm{\gamma}^{\mathrm{rb}})$ \\
1. \textbf{for} $i=1,\cdots,N_s$ \textbf{do}\\
2. \hspace{0.5cm}  $\varphi_i(\bm{X}_*),\mathcal{X}(\cdot;\bm{\gamma}_i)	\leftarrow \frac{\partial\varphi(\bm{X}_*,t)}{\partial t} = \bm{v}(\varphi(\bm{X}_*,t))$ \& $\bm{v} = \argmin_{\bm{v}} \lVert [\varphi(\bm{X}_*)] - [\bm{X}_i] \rVert^2_{W'} + 
    \int_0^1 \lVert\bm{v} \rVert^2_W \textrm{d}t$  \\
3. \hspace{0.5cm} $\bm{\gamma}_i \leftarrow [\bm{e}_1^{\mathrm{T}}(\varphi_i(X_*) - X_*)^{\mathrm{T}},\cdots, \bm{e}_d^{\mathrm{T}}(\varphi_i(X_*) - X_*)^{\mathrm{T}}]^{\mathrm{T}}$\\
4. \hspace{0.5cm} $\mathbf{u}^h(\bm{\xi};\bm{\gamma}_i) \leftarrow a(\bm{w}^h,\bm{u}^h;\bm{\gamma}_i) + c(\bm{u}^h,\bm{w}^h,\bm{u}^h;\bm{\gamma}_i) + b(\bm{w}^h,p^h;\bm{\gamma}_i) = (\bm{w}^h,\bm{\mathrm{g}}^h_{\text{N}})$\\
5.  \hspace{6.95cm}  $b(\bm{u}^h,q^h;\bm{\gamma}_i) = 0$\\
6. \textbf{end for}\\
7. $\{ \bm{\gamma}_i^{\text{rb}} \}_{i=1}^{N_s}, Q \leftarrow \min_{\Tilde{\bm{Q}} \in \mathcal{Z}_{\gamma}} \lVert \bm{\mathrm{M}_{\gamma}} -  \Tilde{Q} \Tilde{Q}^{\mathrm{T}} \bm{\mathrm{M}_{\gamma}} \rVert ^2 _{F}, M_{\gamma}=[ \bm{\gamma}_1|\cdots| \bm{\gamma}_{N_s}] $\\
8. $\{\bm{\mathrm{u}}^{\text{rb}}(\bm{\gamma}^{\mathrm{rb}}_i)\}_{i=1}^{N_s}, \bm{\Phi} \leftarrow \min_{\Tilde{\bm{\Phi}} \in \mathcal{Z}_{\mathrm{u}}} \lVert \bm{\mathrm{M}_{\mathrm{u}}} -  \Tilde{\bm{\Phi}} \Tilde{\bm{\Phi}}^{\mathrm{T}} \bm{\mathrm{M}}_{\mathrm{u}} \rVert ^2 _{F}, M_{\mathrm{u}} = [ \bm{\mathrm{u}}^{h}_1|\cdots| \bm{\mathrm{u}}^{h}_{N_s}] $  \\
9. $\hat{f}(\bm{\gamma}^{\mathrm{rb}}) \leftarrow \bm{\mathrm{u}}^{\mathrm{rb}} \approx \sum_{i=1}^{\ell} \lambda_i \psi(\| \bm{\gamma}^{\mathrm{rb}} -\bm{\gamma}^{\mathrm{rb}}_i \|) + P(\bm{\gamma}^{\mathrm{rb}}),$ 
\end{algorithm}

\begin{algorithm}[b!]
\caption{Online stage of the surrogate model}
\label{alg:PODonline}

\textbf{Input}: Shape of prediction domain $\hat{\bm{X}}=\{\hat{\bm{x}}_{i}\}^{\hat{b}}_{i=1}$; Shape of the reference domain $\bm{X}_*=\{\bm{x}_{*,i}\}^{b_{*}}_{i=1}$;\\
\hspace*{1cm} Reduced bases of geometric parameters $Q$; Reduced bases of flow fields $\bm{\Phi}$; RBF-ROM interpolator, $\hat{f}(\bm{\gamma}^{\mathrm{rb}})$\\
\textbf{Output}: flow field prediction $\hat{\textbf{u}}^h({\mathcal{X}}^{-1}(\cdot;\hat{\bm{\gamma}});\hat{\bm{\gamma}}) $ \\
1. $\hat{\varphi}(\bm{X}_*),{\mathcal{X}}(\cdot;\hat{\bm{\gamma}})	\leftarrow \frac{\partial\varphi(\bm{X}_*,t)}{\partial t} = \bm{v}(\varphi(\bm{X}_*,t))$ \& $\bm{v} = \argmin_{\bm{v}} \lVert [\varphi(\bm{X}_*)] - [\hat{\bm{X}}] \rVert^2_{W'} + 
    \int_0^1 \lVert\bm{v} \rVert^2_W \textrm{d}t$ \\
2. $\hat{\bm{\gamma}} \leftarrow [\bm{e}_1^{\mathrm{T}}(\hat{\varphi}(X_*) - X_*)^{\mathrm{T}},\cdots, \bm{e}_d^{\mathrm{T}}(\hat{\varphi}(X_*) - X_*)^{\mathrm{T}}]^{\mathrm{T}}$\\
3. $\hat{\bm{\gamma}}^{\text{rb}} \leftarrow \hat{\bm{\gamma}} \approx Q\hat{\bm{\gamma}}^{\mathrm{rb}}$\\ 
4. $\hat{\bm{\mathrm{u}}}^{\text{rb}}(\hat{\bm{\gamma}}) \leftarrow \hat{\bm{\mathrm{u}}}^{\text{rb}}(\hat{\bm{\gamma}}^{\text{rb}})=\hat{f}(\hat{\bm{\gamma}}^{\mathrm{rb}}) $\\
5. $\hat{\bm{\mathrm{u}}}^h({\mathcal{X}}^{-1}(\cdot;\hat{\bm{\gamma}});\hat{\bm{\gamma}}) \leftarrow \hat{\bm{\mathrm{u}}}^h(\bm{\xi};\hat{\bm{\gamma}}) = \bm{\Phi}\hat{\bm{\mathrm{u}}}^{\mathrm{rb}}(\hat{\bm{\gamma}})$ \& ${\mathcal{X}}^{-1}(\cdot;\hat{\bm{\gamma}})  $   
\end{algorithm}

In the framework of ROM, there are two stages, online and offline. The offline stage prepares essential ingredients, while the online stage performs predictions based on those ingredients. The offline stage of our proposed method mainly consists of three parts: surface registration, FOM evaluations on the reference domain and the construction of ROM. The details of the offline stage are shown in Algorithm~\ref{alg:PODoffline}.

We are interested in predicting the flow fields on domains of different shapes. Those shapes can be considered as a group and are considered to be similar to each other such that all those shapes can be achieved by some small deformations from a reference domain. For a computational model simulating e.g. time-dependent tissue growth in the artery, the initial shape of the domain can be chosen as the reference shape. For more general situations, the reference shape can be chosen as the average of the shapes, namely the atlas construction \cite{Durrleman2012,Gori2013}. Here, we denote those shapes used in the offline stage as our training data and can be presented by a collection of boundary points $\{\bm{X}_j\}_{j=1}^{N_s} = \{\{\bm{x}_{1,i}\}^{b_{1}}_{i=1},\{\bm{x}_{2,i}\}^{b_{2}}_{i=1},\cdots,\{\bm{x}_{{N_s},i}\}^{b_{N_s}}_{i=1} \}$, where $b_i$ denotes the number of boundary points for each shape and $N_s$ denotes the total number of shapes available for training. 
The reference shape can be represented in a similar way, $\bm{X}_*=\{\bm{x}_{*,i}\}^{b_{*}}_{i=1}$. Those boundary points will serve as discrete CPs used in the kernel computation during the registration process. In practice, the vertices/nodes on the boundary of a mesh prepared for traditional numerical methods, e.g. FEM, can be directly considered as the CPs.

\begin{figure}[tb!]
\centering
 \includegraphics[width=1.0\textwidth]{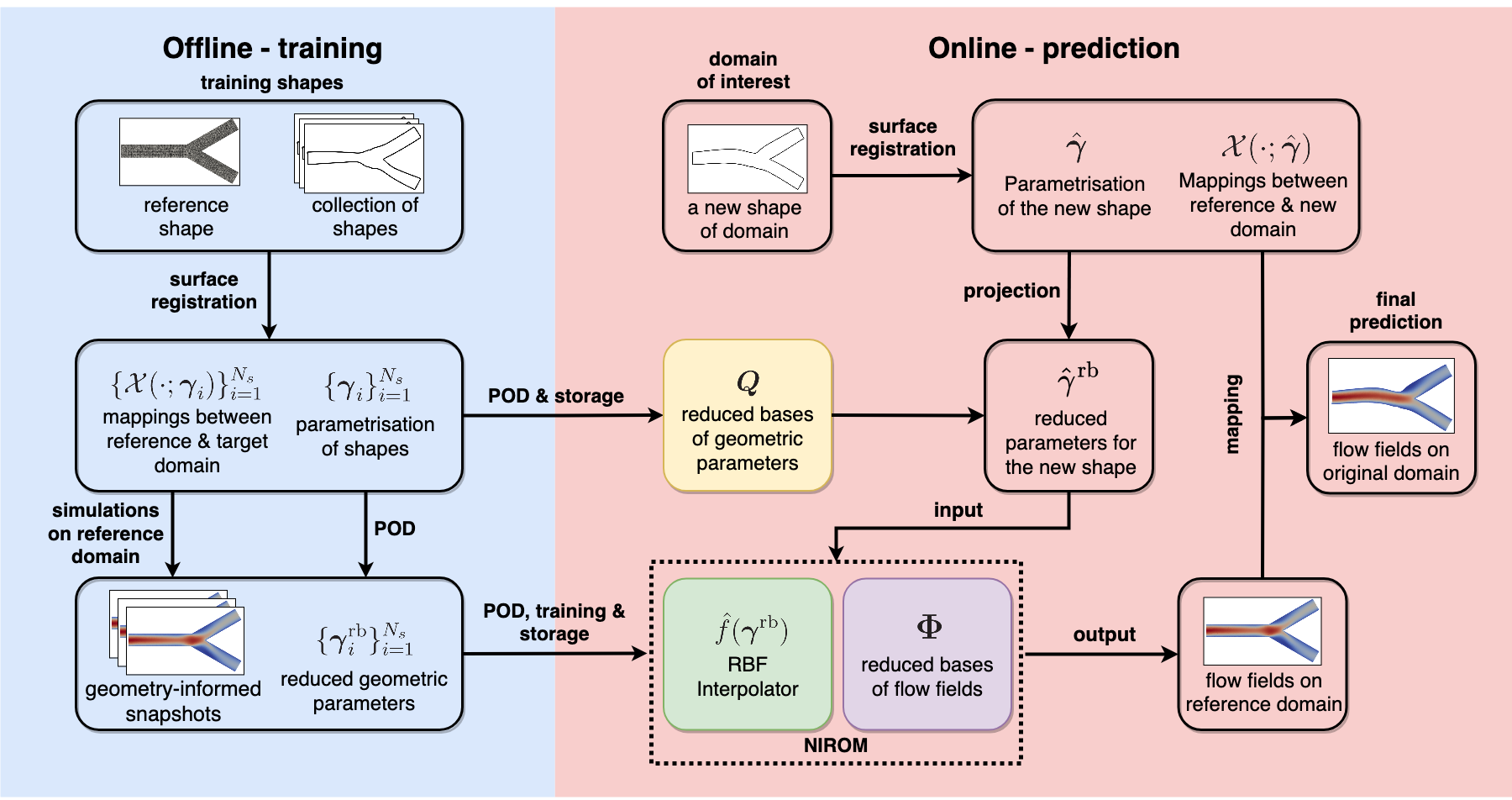}
 \caption{A schematic diagram of the proposed surrogate model for blood flow simulations. POD, proper orthogonal decomposition; NIROM, non-intrusive reduced-order model; RBF, radial basis function} 
\label{fig:ROMSR}
\end{figure}

By performing the surface registration, (equation~\ref{eq:flowofdiffeo} and equation~\ref{eq:velocityoptimzation}) based on currents, the displacements of the CPs from the reference shape associated to its particular diffeomorphisms $\varphi_i$ are computed and viewed as the geometric parameters of the shape, i.e. $\bm{\gamma}_i=[\bm{e}_1^{\mathrm{T}}(\varphi_i(X_*) - X_*)^{\mathrm{T}},\cdots, \bm{e}_d^{\mathrm{T}}(\varphi_i(X_*) - X_*)^{\mathrm{T}}]^{\mathrm{T}}$, for $i=1,\cdots,N_s$. As mentioned before, such a parametrisation method may result in exceedingly high-dimensional parameters. Therefore POD is employed to extract their reduced representations $\{\bm{\gamma}^{\text{rb}}_i\}_{i=1}^{N_s}$ based on equation~\ref{eq:reducedgeoparameter}. Note that this process is similar to parameter identification in statistical shape analysis \cite{zheng2017statistical}, where the POD is commonly utilised for the parametrisation of a group of shapes with landmark correspondence.

On the other hand, the complete diffeomorphism between the reference domain and an arbitrary domain encompassing a shape from the training dataset can be approximated by RBF interpolation based on the positions of the CPs before and after the deformation (registration). The approximated mapping, denoted as $\mathcal{X}(\cdot;\bm{\gamma}_i) \approx \varphi_i(\cdot)$, is subsequently employed to project the Navier-Stokes equations onto the reference domain using the Jacobian matrix $J_{\mathcal{X}_i}$ of each $\mathcal{X}(\cdot;\bm{\gamma}_i)$ and its determinant $|J_{\mathcal{X}_i}|$ for $i=1,2,\cdots,N_s$. As a result, the Navier-Stokes equations with different shapes of domains can all be resolved on the reference domain $\Omega^*$ and the corresponding discrete geometry-informed snapshot $\{\mathbf{u}^h(\bm{\xi};\bm{\gamma}_i)\}^{N_s}_{i=1}$ can be generated (equation~\ref{eq:snapshotgeneration}). This approach ensures spatial compatibility over the snapshots corresponding to the geometric parameters for ROM.

The final step in the offline stage is to construct the ROM. The orthogonal reduced bases $\bm{\Phi}$ of the parametric ROM are computed by applying POD to the snapshot matrix $[\mathbf{u}^h(\mathbf{\gamma}_1)|\cdots|\mathbf{u}^h(\bm{\gamma}_{N_s})]$ (equation~\ref{eq:orthbasis}). $N_l$ number of the bases is selected to ensure the cumulative energy $\epsilon$ of the bases will cover $99.9\%$ of the total energy. The corresponding reduced coefficients $\mathbf{u}^{\text{rb}}(\bm{\gamma}_j)$, where $j=1,\cdots,N_s$, are computed via projecting the snapshots onto low-rank space. The mapping between reduced coefficients of the flow field $\{\mathbf{u}^{\text{rb}}(\bm{\gamma}^{\text{rb}}_i)\}^{N_s}_{i=1}$ and the reduced geometrical parameters $\{\bm{\gamma}^{\text{rb}}_i\}^{N_s}_{i=1}$ is also approximated by RBF interpolation (equation~\ref{eq:chap4_RBFI}).

During the online stage, ROM performs the flow dynamics prediction associated with the domain of a new geometry, as shown in Algorithm~\ref{alg:PODonline}. Similarly to the training procedure, the new geometry will be parameterised and compressed through registration and dimensionality reduction, respectively. The reduced geometric parameters are subsequently fed to the trained RBF interpolator to estimate the reduced coefficients of the corresponding flow fields. The estimated reduced coefficients, together with the reduced bases derived in the offline stage, reconstruct the entire flow field on the reference domain, which is mapped back to its original domain in the final step. Figure~\ref{fig:ROMSR} illustrates the comprehensive workflow of the proposed method.

\subsection{Error estimation}
In order to present the accuracy and performance of the proposed ROM, several measurements were applied. The relative error was employed to visualise the relative difference between two flow fields at a specific point $\bm{x}$,
\begin{equation}
    e(\bm{x}) = \frac{|\textbf{u}^h(\bm{x})-\hat{\textbf{u}}^h(\bm{x})|}{max\{|\textbf{u}^h|\}}.
\end{equation}
where $max\{|\textbf{u}^h|\}$ denotes the maximum absolute value of the vector $\textbf{u}^h$. Relative $L^2$ error corresponding to geometric parameters $\bm{\gamma}$ is employed to measure the overall difference of the estimations between a FOM and the proposed ROM over the domain,

\begin{equation}
    \epsilon_{\text{rb}}(\bm{\gamma}) = \frac{ \|\textbf{u}^h(\bm{\gamma}) - \hat{\textbf{u}}^h(\bm{\gamma}) \|_{L^2}}{\| \textbf{u}^h(\bm{\gamma}) \|_{L^2}} = \frac{ \|\textbf{u}^h(\bm{\gamma}) - \bm{{\Phi}}\textbf{u}^{\text{rb}}(\bm{\gamma}) \|_{L^2}}{\| \textbf{u}^h(\bm{\gamma}) \|_{L^2}}. 
\end{equation}
where $\|\cdot\|_{L^2}$ denotes $L^2$ norm. Besides, the projection error introduced by POD is also measured by the relative $L^2$ error,
\begin{equation}
    \epsilon_{\text{POD}}(\bm{\gamma})= \frac{ \|\textbf{u}^h(\bm{\gamma}) - \bm{{\Phi}} \bm{{\Phi}}^{\mathrm{T}} \textbf{u}^h(\bm{\gamma})\|_{L^2}}{\| \textbf{u}^h(\bm{\gamma}) \|_{L^2}}. 
\end{equation}

\section{Numerical examples}\label{sec:example}
\subsection{Case one: blood flow in a stenosis}
\begin{figure}[tb]
\centering
 \includegraphics[width=1.0\textwidth]{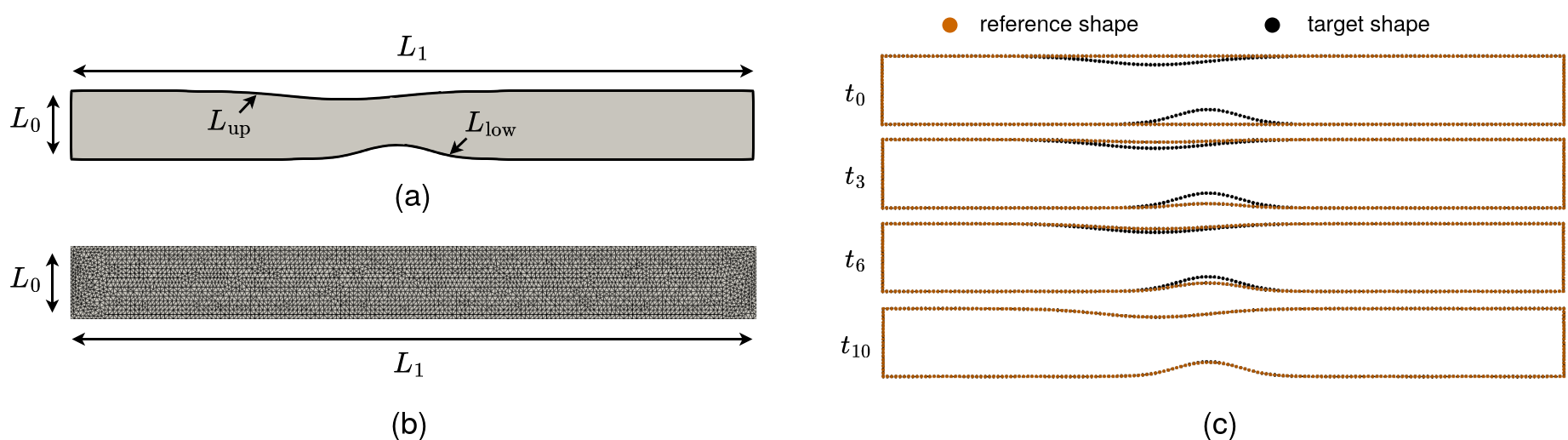}
 \caption{(a) Geometric configuration of the stenosis domain. $L_0 = 2 \,$mm and $L_1=20\,$mm denote the height and width of the domains. $L_{\text{low}}$ and $L_{\text{up}}$ denote Gaussian function for upper and lower boundaries. (b) The reference mesh used for surface registration and FEM implementation. (c) One example of surface registration results. The reference shape deforms gradually toward the target shape using the flow of diffeomorphism. The shapes are represented by a series of boundary points.}
\label{fig:shape_of_stenosis}
\end{figure}

\begin{figure}[tb]
\centering
\hspace{1.5cm}
 \includegraphics[width=0.56\textwidth]{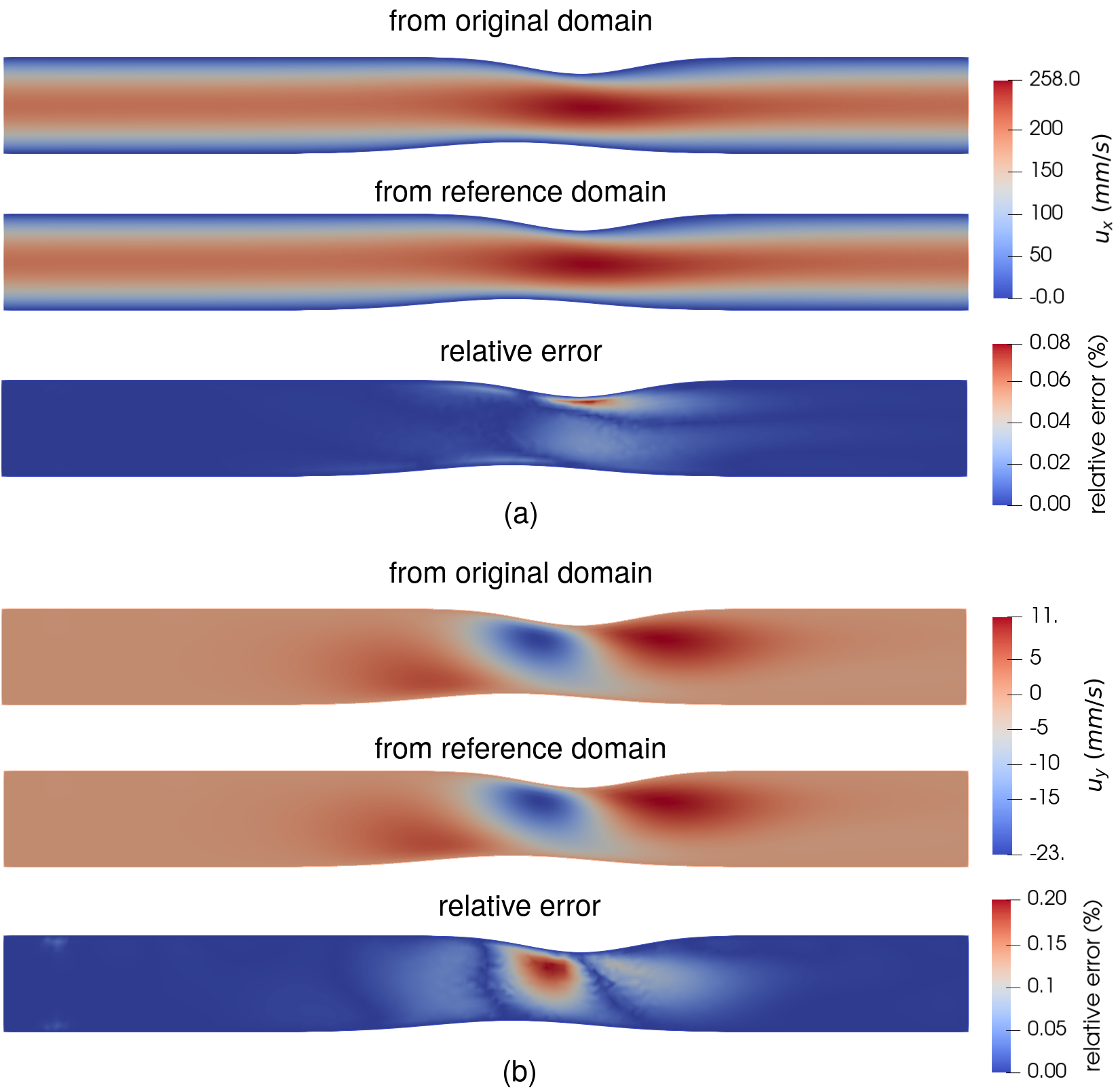}
 \caption{A comparison of the flow fields solved on the reference domain and its original domain. a) Velocity fields and corresponding relative error in the x direction. b) Velocity field and corresponding relative error in the y direction.}
\label{fig:reference_steady}
\end{figure}

\begin{figure}[h!]
\centering
 \includegraphics[width=1\textwidth]{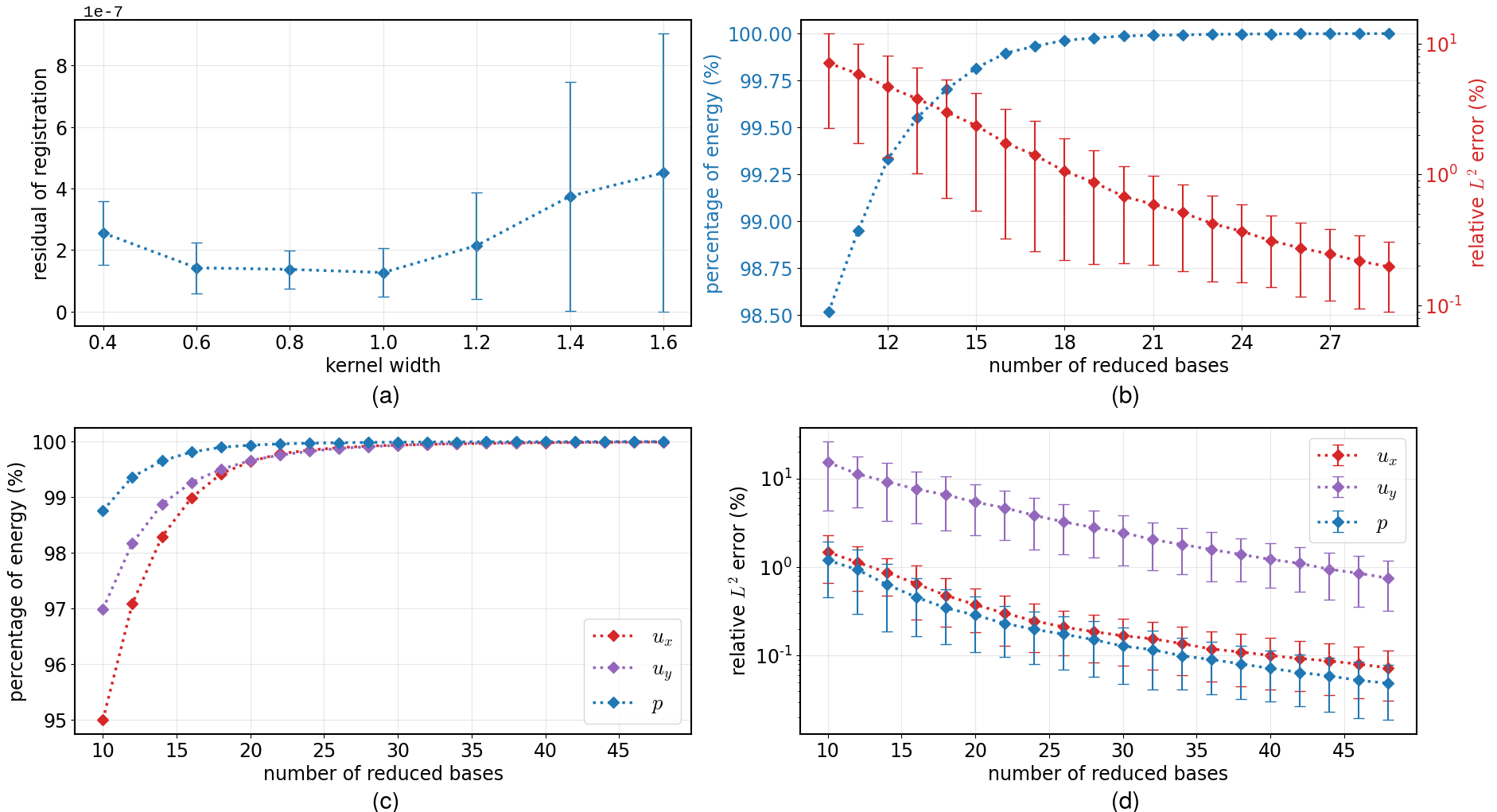}
 \caption{ (a) Mean and standard deviation of the residual of surface registrations with different kernel widths $\lambda_W$ for stenosis geometries. (b) Percentage of energy captured by the number of the reduced bases of geometric parameters and its corresponding relative $L^2$ error. (c)-(d) Percentage of energy captured by the number of the reduced bases of the velocity fields and pressure fields and their corresponding relative $L^2$ error.} 
\label{fig:chap4POD}
\end{figure}

\begin{figure}[t!]
\centering
 \includegraphics[width=1\textwidth]{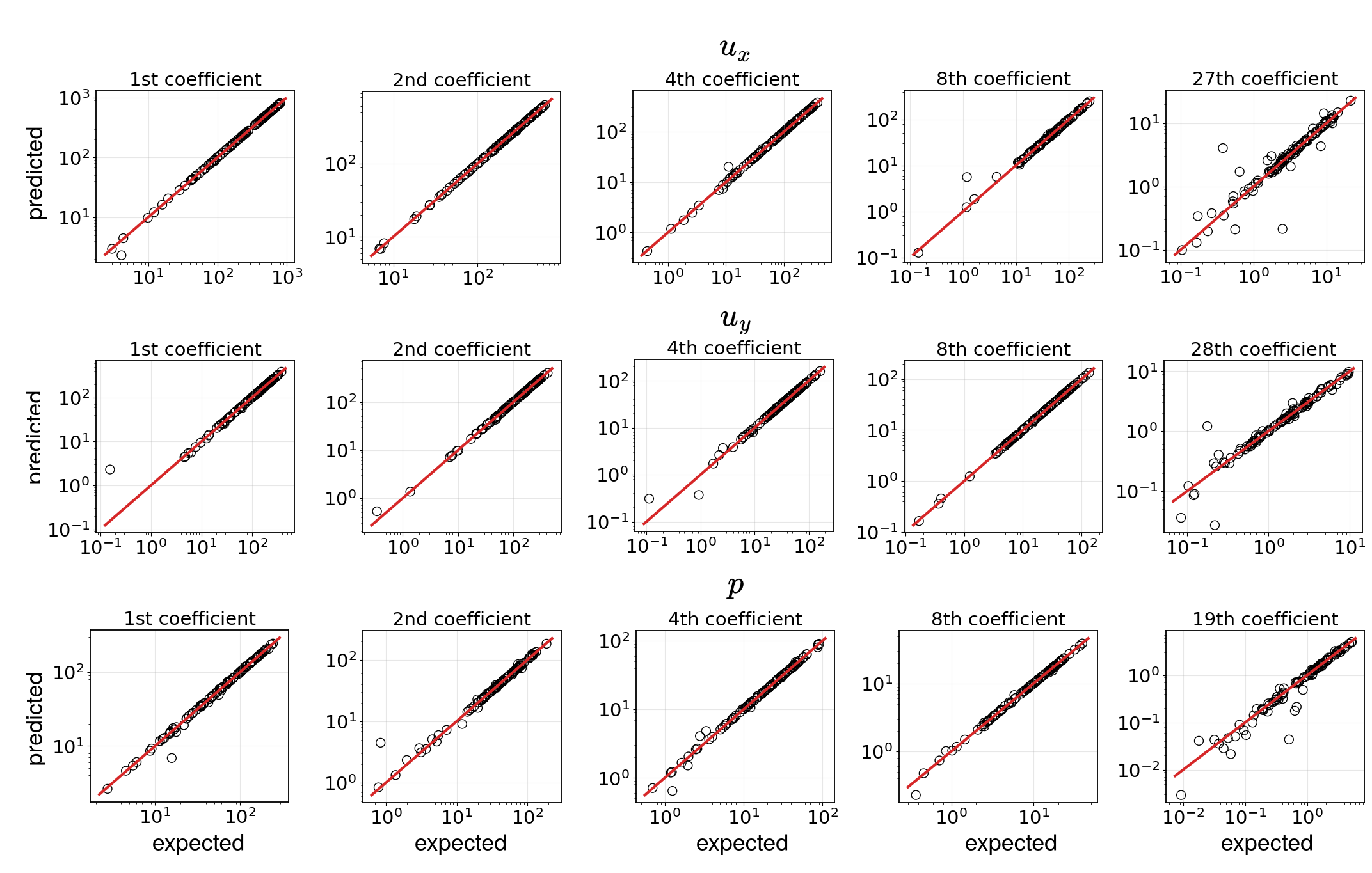}
 \caption{A comparison between expected and predicted 1st, 2nd, 4th, 8th and last reduced coefficients of velocity in x and y directions and pressure (absolute values). The predictions of reduced coefficients are provided by trained RBF interpolator in ROM. The circles represent the ROM prediction of reduced coefficients by the RBF interpolator and the diagonal lines denote the precise predictions of expected values.}
\label{fig:chap4RC}
\end{figure}

\begin{figure}[tb]
\centering
 \includegraphics[width=0.85\textwidth]{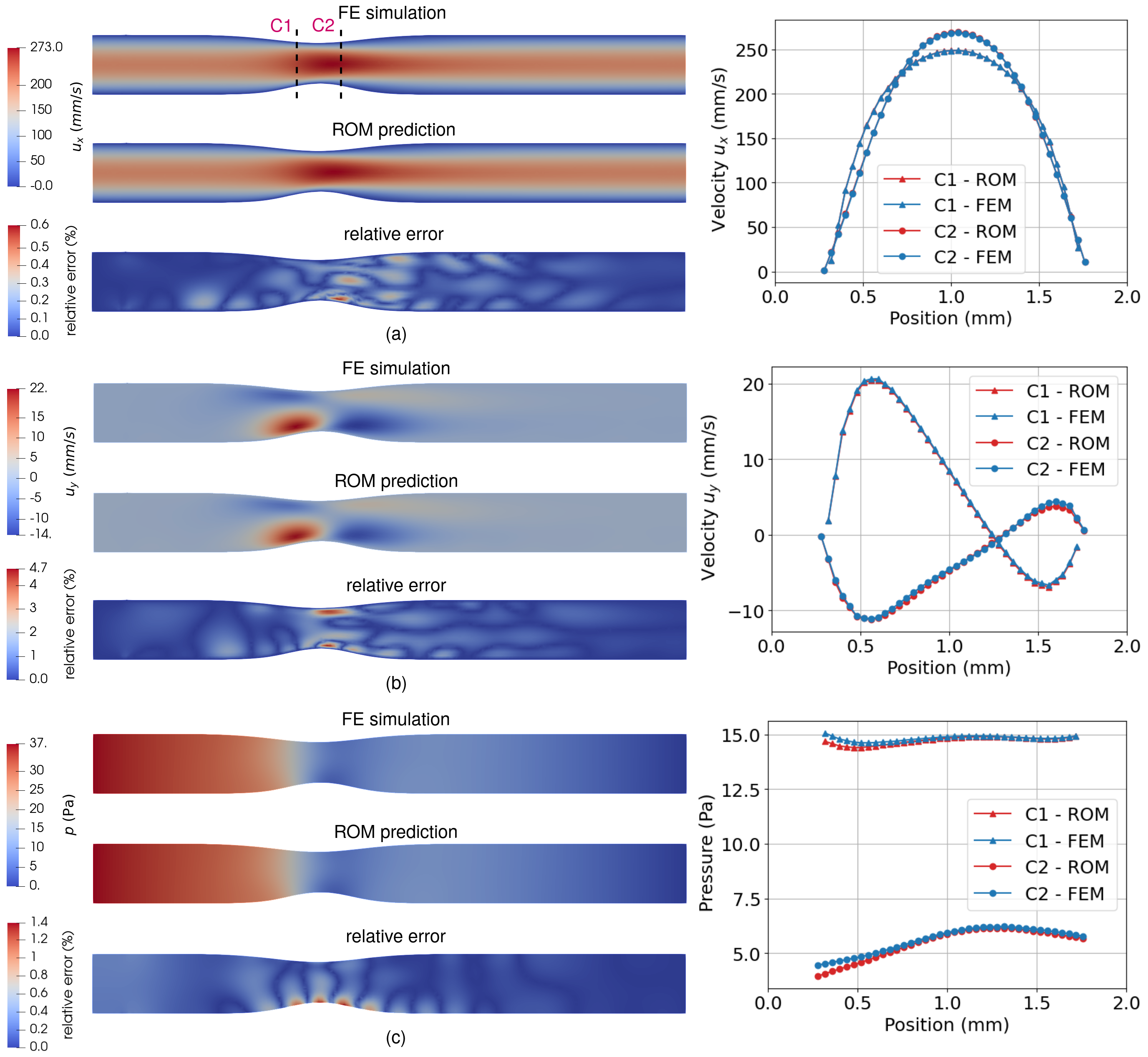}
 \caption{A comparison between FE simulation results and ROM prediction for velocities and pressure and their corresponding relative errors. A detailed comparison of three properties on cross-lines C1 and C2 is presented on the left.} 
\label{fig:chap4steady_u}
\end{figure}

The simulation of blood flowing through a stenotic vessel is a common scenario in cardiovascular science to study the development of arteriosclerosis or in-stent restenosis. We constructed an idealised 2D example of blood flow passing through a stenotic coronary artery. The geometric configuration of the domain is shown in Figure~\ref{fig:shape_of_stenosis}(a). The position and severity of the stenosis of the upper and lower boundaries are described by the means and standard deviations of two Gaussian functions,
\begin{equation*}
    L_{\text{low}}(x) = \frac{1}{\sqrt{2\pi\sigma_1^2}}\textrm{exp}\Big(-\frac{(x-\mu_1)^2}{2\sigma_1^2}\Big),
\end{equation*}
\begin{equation*}
    L_{\text{up}}(x) = L_0 -  \frac{1}{\sqrt{2\pi\sigma_2^2}}\textrm{exp}\Big(-\frac{(x-\mu_2)^2}{2\sigma_2^2}\Big),
\end{equation*}
where $L_0=2 \,\text{mm}$ is the width of the inlet. 500 different shapes of domains were generated by sampling $\sigma_1$ and $\sigma_2$ within the ranges $[0.8,2]$, $\mu_1$ and $\mu_2$ within the range $[0.3L_1,0.7L_1]$ using the Latin hypercube method \cite{Michael1987}. 400 of the samples were applied for the construction of the ROM in the offline stage, while the remaining 100 samples were equally split into the validation dataset and test dataset. The dynamic viscosity and density of the flow were set to be $3.5\times10^{-3}\, \text{g}/\text{(mm}\cdot\text{s)}$ and $1.06\times 10^{-3}\, \text{g}/\text{mm}^3$. A constant parabolic velocity profile with a maximum velocity of $200 \text{mm/s}$ was prescribed at the inlet. A non-slip boundary condition was prescribed for the upper and lower boundaries, and a homogeneous Neumann condition was forced on the outlet. The simulation including mesh generation and finite element method was implemented using the open-source PDE solver \textit{FreeFEM} \cite{MR3043640}. A rectangular shape was chosen for the shape of the reference domain, as shown in Figure~\ref{fig:shape_of_stenosis}(b) and a Taylor-Hood P2-P1 finite element spatial discretisation was applied.

\begin{table}[th!]
\centering
\begin{adjustbox}{width=1\textwidth}
\begin{tabular}{c|c|c|c|c|c|c}
\toprule
Method & Estimated Property & \begin{tabular}[c]{@{}c@{}}Dimension \\of FOM/ROM\end{tabular}  & \begin{tabular}[c]{@{}c@{}}Relative $L^2$ error \\ $\pm$ SD (POD)\end{tabular} & \begin{tabular}[c]{@{}c@{}}Relative $L^2$ error \\ $\pm$ SD (Surrogate)\end{tabular} &\begin{tabular}[c]{@{}c@{}}Computational time\\ of prediction - FOM/ROM (s)\end{tabular} & \begin{tabular}[c]{@{}c@{}}Computational time\\ of prediction - SR (s)\end{tabular}   \\
\midrule
FOM    & $u_x,u_y,p$        & 2991     & / &  /    &      4.39    & /        \\
\cmidrule(lr){1-7}
ROM+SR & $u_x$       & 27  & $0.19 \pm 0.11 \%$  &   $0.23 \pm 0.16 \%$     &   8.21 $\times 10^{-5}$    &     \multirow{3}{*}{1.03}          \\
\cmidrule(lr){1-6}
ROM+SR &  $u_y$        & 28   & $2.80 \pm 1.52 \%$   &   $3.27 \pm 2.23\%$     & 9.16 $\times 10^{-5}$  &          \\
\cmidrule(lr){1-6}
ROM+SR &  $p$        & 19   &  $0.32 \pm 0.21 \%$  &  $0.48\pm 0.50\%$     & 6.89 $\times 10^{-5}$  &      \\
\bottomrule
\end{tabular}
\end{adjustbox}
\caption{Error estimation and computational time of the FOM and proposed surrogate model based on ROM with geometry-informed snapshots. The relative $L^2$ error of POD and the entire surrogate model and the computational time were evaluated based on the test dataset with 50 samples. The computational time of prediction consists of two parts: the average computational time of surface registrations (SR) and non-intrusive reduced order model (ROM) evaluations. Note that the FOM here denotes performing a simulation with FOM on its original domain. All the predictions here were performed on Intel Core i7-7500U CPU@2.70GHz.}
\label{tab:speedup_case1}
\end{table}

% \usepackage{multirow}
{\renewcommand{\arraystretch}{1.2}
\begin{table}[th!]
\centering
\begin{adjustbox}{width=1\textwidth}
\begin{tabular}{ccc|cccc|c}
\toprule
\multicolumn{3}{c|}{Registration} 
& \multicolumn{4}{c|}{ROM}  
& Other  \\ \hline
\multicolumn{1}{c|}{\begin{tabular}[c]{@{}c@{}}surface \\ registration\end{tabular}} 
& \multicolumn{1}{c|}{\begin{tabular}[c]{@{}c@{}}geometric parameters \\ compression (SVD)\end{tabular}} 
& \begin{tabular}[c]{@{}c@{}}diffeomorphism \\ appproximation (RBFI)\end{tabular} 
& \multicolumn{1}{c|}{properties} 
& \multicolumn{1}{c|}{\begin{tabular}[c]{@{}c@{}}reduced basis \\ construction (SVD)\end{tabular}} 
& \multicolumn{1}{c|}{\begin{tabular}[c]{@{}c@{}}reduced coefficient \\ mapping (RBFI)\end{tabular}} 
& \begin{tabular}[c]{@{}c@{}}Snapshot \\ generation\end{tabular} 
& Data processing  \\ \hline

\multicolumn{1}{c|}{\multirow{3}{*}{$1.03N_s $}}  
& \multicolumn{1}{c|}{\multirow{3}{*}{0.25}}   
& \multirow{3}{*}{0.041}                                                               
& \multicolumn{1}{c|}{$u_x$}      
& \multicolumn{1}{c|}{$0.36 $}                                                        
& \multicolumn{1}{c|}{$0.085 $}                                                         
& \multirow{3}{*}{12.2$N_s$}                                              
& \multirow{3}{*}{68.8 } \\ \cline{4-6}
\multicolumn{1}{c|}{}                                                                
& \multicolumn{1}{c|}{}                                                                                  &                                                                                
& \multicolumn{1}{c|}{$u_y$}      
& \multicolumn{1}{c|}{$0.45 $}                                                        
& \multicolumn{1}{c|}{$0.019 $}                                                         
&                                                               
&                  \\ \cline{4-6}
\multicolumn{1}{c|}{}                                                               
& \multicolumn{1}{c|}{}                                                                                  &                                                                                 
& \multicolumn{1}{c|}{$p$}        
& \multicolumn{1}{c|}{$0.42 $}                                                        
& \multicolumn{1}{c|}{$0.015$}                                                         
&                                                                
&                   \\ \bottomrule
\end{tabular}
\end{adjustbox}
\caption{Computational time (sec) of offline components of the proposed method for the stenosis case. $N_s =400$ denotes the number of data used for training. All the computations here were performed on Intel Core i7-7500U CPU@2.70GHz, except that the snapshot generation was carried out on a shared AMD Rome 7H12 node with 16-core on Dutch supercomputer Snellius.}
\label{tab:offline1}
\end{table}}

The surface registration was performed via statistical analysis software \textit{Deformetrica} \cite{deformatrica,DURRLEMAN201435}. The residual of registration based on different choices of kernel width was shown in Figure~\ref{fig:chap4POD}(a). The minimum value of the residual mean was achieved when the kernel width is set to one and the standard deviation was relatively small. One of the surface registration results is demonstrated in Figure~\ref{fig:shape_of_stenosis}(c). The result shows that the reference shape deformed gradually toward the target shape and aligned well at the final time step. The average residual of the surface registration over 500 samples is $1.31\times10^{-7}$(measured on currents). The results of surface registration were subsequently applied to compute the mappings between the domains of different shapes using RBF interpolation. A cubic kernel is employed with a polynomial function of the first order. Owing to the limited amount of mapping information we have, it is difficult to validate the interpolation itself. Instead, the validation was performed based on the results of the FEM simulation. The result computed on the reference domain was mapped back to its original domain and compared to the result that was directly simulated on the target domain. The accuracy of the simulation result based on the reference domain reflects the accuracy of the mapping. Figure~\ref{fig:reference_steady} shows the comparison of one selected sample. The regions with relatively higher errors are located close to the deformed boundaries, which means that RBF interpolation for the mapping introduced minor errors into the snapshots. The average relative $L^2$ errors of velocities in x and y directions, and pressure of 400 training samples are 0.21\%, 0.42\% and 0.042\%, respectively. It implies that the overall errors introduced by mapping are small, therefore the mapping approximated by the RBF interpolation is accurate enough for the FE simulation. We therefore considered these FOM evaluations based on the reference domain as the snapshots for the reduced model. 

The surface registration results have also been applied to parameterise the configuration of the domains. In this case, there are 360 points on the boundary and therefore leads to 720 geometric parameters in a two-dimensional problem. POD was applied to further reduce the dimensionality of the parameters to 17, which has already captured $99.9\%$ of the total variance. The percentage of the total variance captured by the number of reduced bases and their corresponding $L^2$ error of projection is demonstrated in Figure~\ref{fig:chap4POD}(b). The error decreased almost linearly on the log scale.

The discrete results of simulations on the reference domain are subsequently taken as the snapshots of ROM, and POD was performed to build reduced bases. The percentage of the total variance of the velocity and pressure data captured by the number of reduced bases is demonstrated in Figure~\ref{fig:chap4POD}(c). Similar to the dimensionality reduction to the geometric parameters, we assume that if more than $99.9\%$ of the variance has been recovered by POD, the approximation reconstructed by the first $N_l$ bases performs well enough. More rigorous criteria could be set but the benefits are marginal as long as the main patterns have been already covered. The relative $L^2$ errors introduced by POD are visualised in Figure~\ref{fig:chap4POD}(d). The relative $L^2$ error of $u_y$ is slightly higher than the other two properties. This may due to that reduced basis constructed from POD is a linear projection of the solution manifold approximated by the snapshots. Therefore, the error can be further diminished by increasing the number of training snapshots or applying nonlinear projection.

RBF interpolation again was utilised to predict reduced coefficients of flow fields based on reduced coefficients of the geometric parameters. A comparison of the expected and predicted reduced coefficients of one test case is shown in Figure~\ref{fig:chap4RC}. The points of the most coefficients cluster around the diagonal line, indicating that the RBF interpolation predicted the reduced coefficients well. Obvious deviations can be found in the last coefficients. This is mainly due to that the latent functions of the coefficients of high-frequency are generally less smooth and more difficult to interpolate. Figure~\ref{fig:chap4steady_u} visualises the velocity fields reconstructed from predicted reduced coefficients and compares the result to the evaluation of the corresponding FOM. Note that in this comparison, the evaluations of the FOM were based on the reference domain. The average relative $L^2$ errors over 50 validation samples are $0.24\%$, $3.42\%$ and $0.47\%$ for velocity in x and y directions and pressure, respectively. The error estimations on the test dataset are shown in Table~\ref{tab:speedup_case1}. The result shows that the prediction of the surrogate model matches the finite element simulation result well. The relative $L^2$ error is comparatively large in the y direction owing to its small magnitude compared to the velocity in the x direction. Note that the error estimations here do not include the error introduced by the construction of the diffeomorphism, which was neglected as they were small enough. They only represent how well the non-intrusive ROM predicted based on given snapshot data.

A detailed comparison of FOM and proposed ROM with geometry-informed snapshots on the computational time and error estimations are listed in Table~\ref{tab:speedup_case1}. It is obvious that the computation of ROM barely took any time, while most of the computational effort of the proposed surrogate model was spent on surface registration. Therefore the entire speedup of the proposed method heavily relies on the performance of surface registration. The error of the ROM mainly originated from POD due to the limitation of the amount of training data. The computational time for each offline component is presented in Table~\ref{tab:offline1}. The major computational cost for the offline stages stemmed from the snapshots generation and took 12.2 sec per evaluation on average. The additional computational cost compared to a direct simulation on its original domain arose from the derivative computation of projection. The data processing includes the time of read/save data between \textit{FreeFEM} and ROM model.

\begin{figure}[tb!]
\centering
 \includegraphics[width=0.9\textwidth]{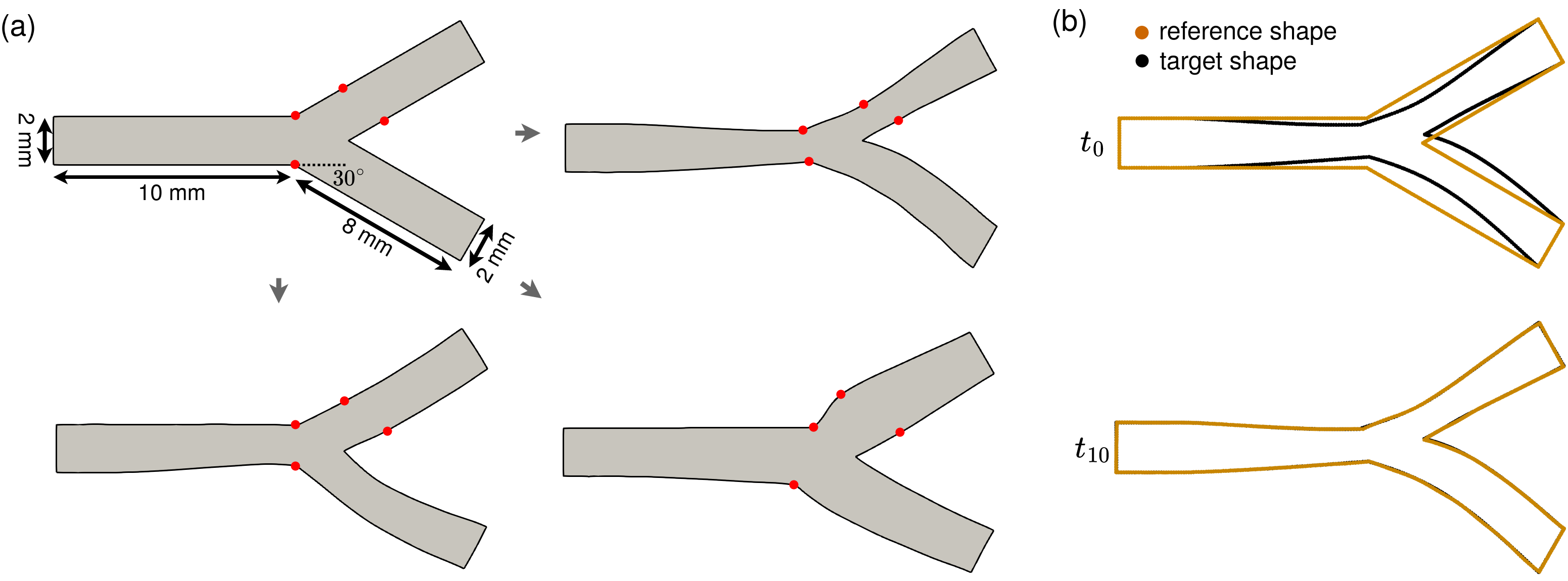}
 \caption{(a) Basic configuration of the bifurcation domain and different shapes generated using four control points (in red) via interpolation. (b) Surface registration results. The reference shape deforms towards the target shape using the flow of diffeomorphism. $t_{10}$ denotes the final registration result.} 
\label{fig:chap4bifur}
\end{figure}

\begin{figure}[th!]
\centering
 \includegraphics[width=1\textwidth]{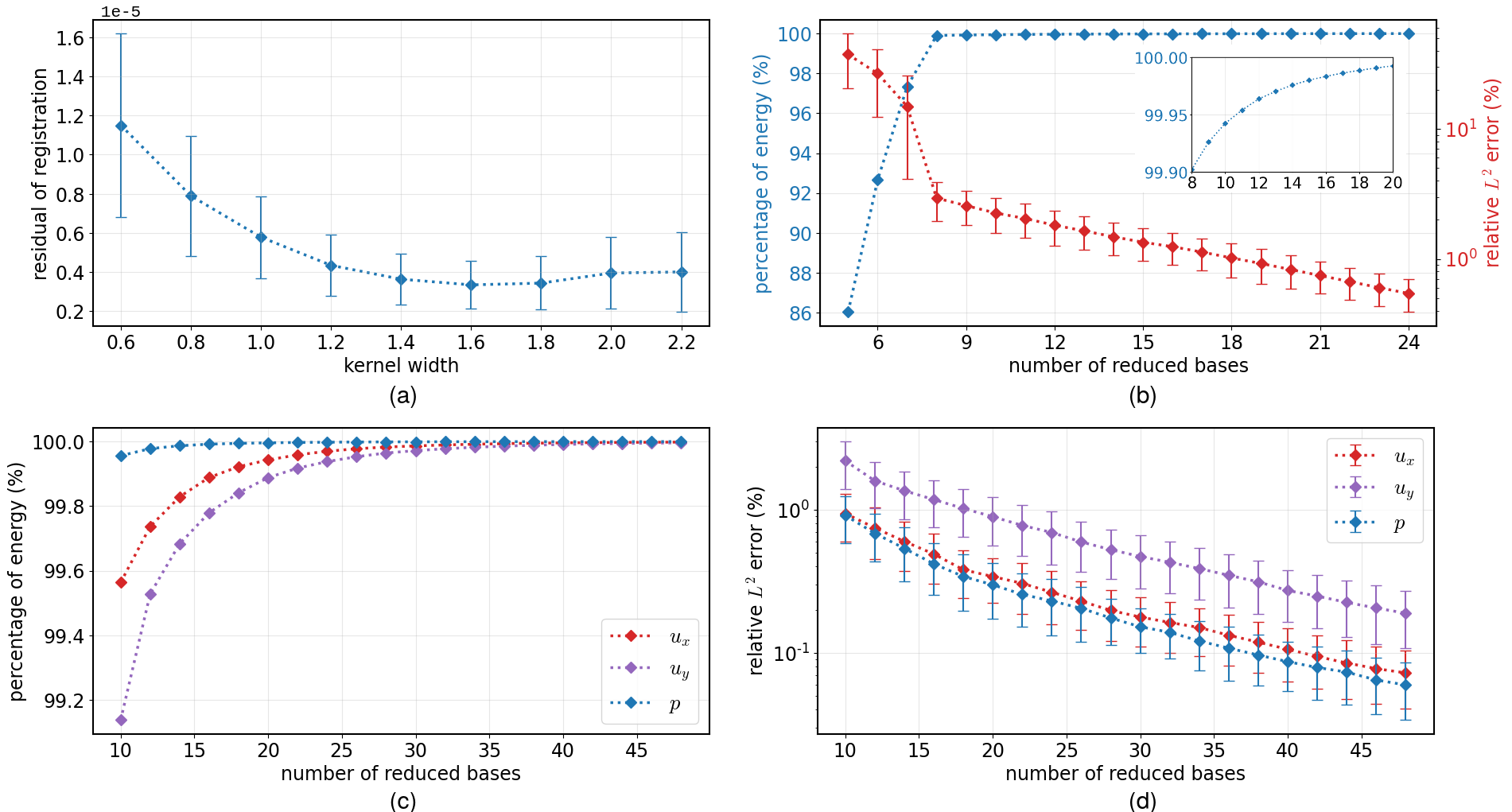}
 \caption{ (a) Mean and standard deviation of the residual of surface registrations with different kernel widths $\lambda_W$ for bifurcation geometries. (b) Percentage of energy captured by the number of the reduced bases of geometric parameters and its corresponding relative $L^2$ error. (c)-(d) Percentage of energy captured by the number of the reduced bases of the velocity fields and pressure fields and their corresponding relative $L^2$ error.} 
\label{fig:BifurPOD}
\end{figure}

\begin{figure}[tb!]
\centering
 \includegraphics[width=0.97\textwidth]{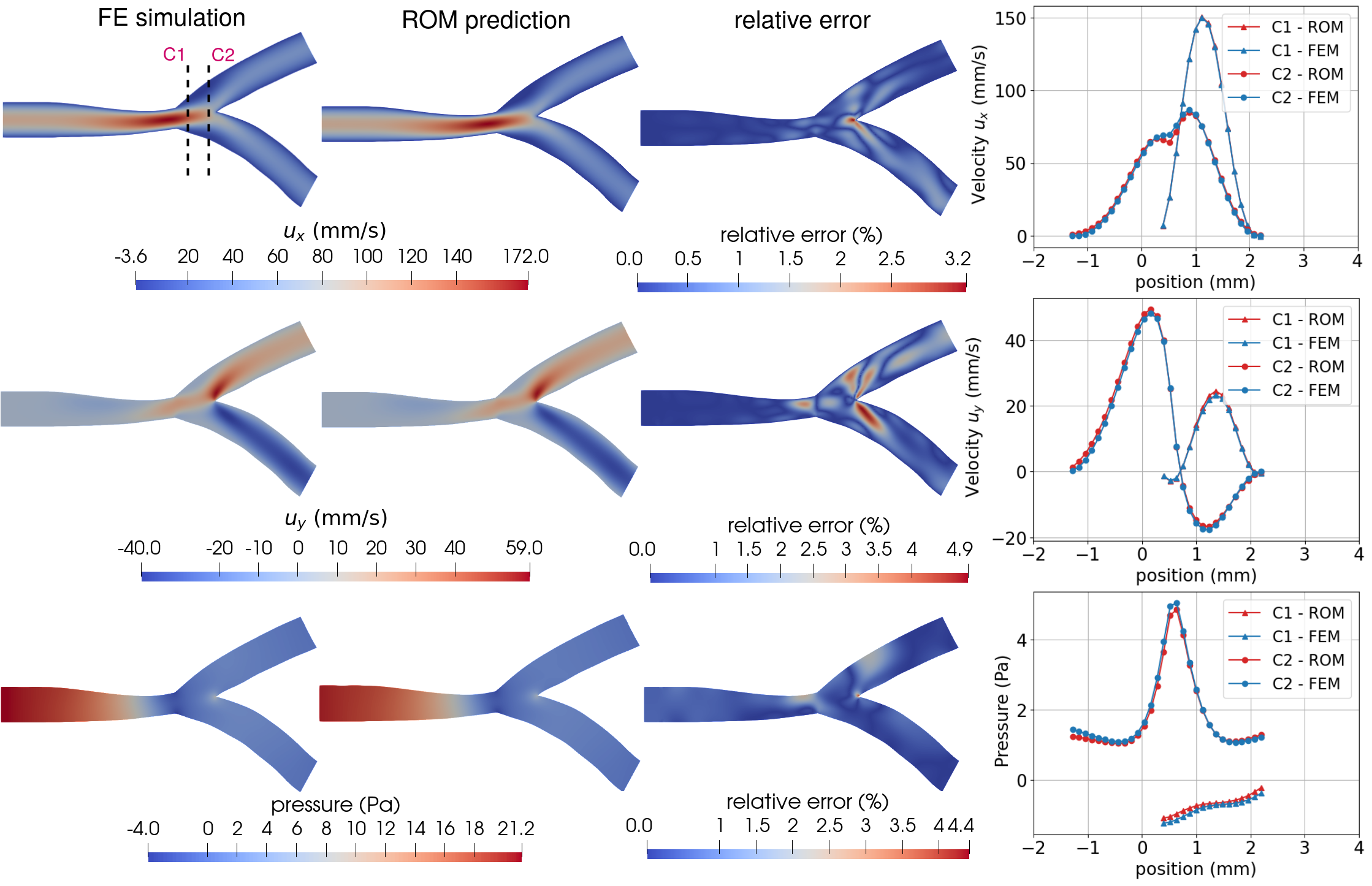}
 \caption{A comparison between FE simulation results and ROM prediction for velocities and pressure and their corresponding relative errors. A detailed comparison of three properties on cross-lines C1 and C2 are presented on the left.} 
\label{fig:chap4bifur_pred}
\end{figure}

\begin{figure}[tb!]
\centering
 \includegraphics[width=0.97\textwidth]{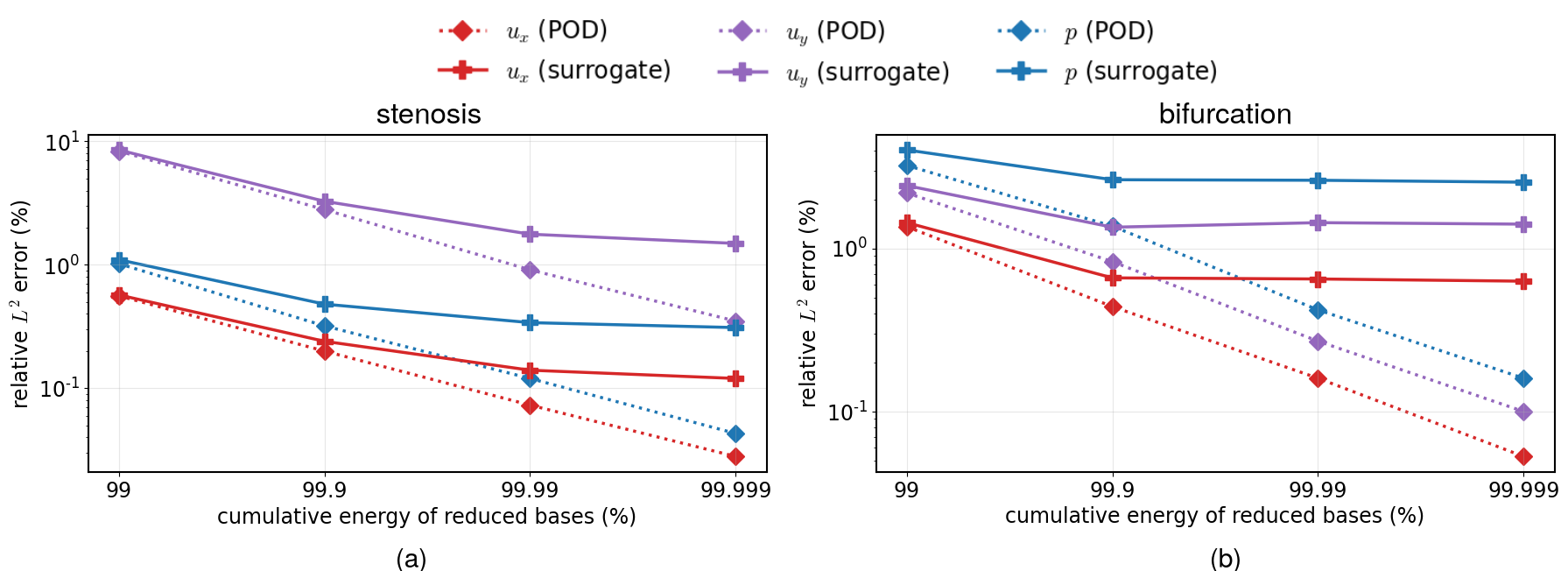}
 \caption{ Relative $L^2$ error of POD and entire surrogate model with reduced bases of different cumulative energy in stenosis (a) and bifurcation (b) cases, respectively. } 
\label{fig:PODincrease}
\end{figure}

% \begin{figure}[tb!]
% \centering
%  \includegraphics[width=0.7\textwidth]{Figure/Bifurcation_pred_detail.png}
%  \caption{A detailed comparison between FE simulation results and ROM prediction on the cross-line C1 and C2 for velocities and pressure and their corresponding relative errors.} 
% \label{fig:chap4bifur_pred_detail}
% \end{figure}

\subsection{Case two: blood flow in a bifurcation}
We present the second example of blood flow simulations through an artery bifurcation, another common scenario in cardiovascular science. The basic configuration of the flow domain is shown in Figure~\ref{fig:chap4bifur}(a), which is also considered as the reference shape. A series of new shapes are generated via deformations based on the basic configuration. The four red dots in Figure~\ref{fig:chap4bifur} are the chosen points to control the deformation. We consider the displacement of these points in the x and y directions as random variables and sample within the range $[-0.5,0.5]$(mm). The new shapes are therefore imposed by interpolation. Similar to case one, 500 samples were generated by the Latin hypercube method and the same fluid properties and boundary conditions were applied. The Reynolds number for the case is 60. 

\begin{table}[tb!]
\centering
\begin{adjustbox}{width=1\textwidth}
\begin{tabular}{c|c|c|c|c|c|c}
\toprule
Method & Estimated Property & \begin{tabular}[c]{@{}c@{}}Dimension \\of FOM/ROM\end{tabular}  & \begin{tabular}[c]{@{}c@{}}Relative $L^2$ error \\ $\pm$ SD (POD)\end{tabular} &\begin{tabular}[c]{@{}c@{}}Relative $L^2$ error \\ $\pm$ SD (Surrogate)\end{tabular} &\begin{tabular}[c]{@{}c@{}}Computational time of\\ prediction - FOM/ROM (s)\end{tabular} & \begin{tabular}[c]{@{}c@{}}Computational time\\ of prediction - SR (s)\end{tabular}   \\
\midrule
FOM    & $u_x,u_y,p$        & 6686     & /  &  /    &      13.22    & /        \\
\cmidrule(lr){1-7}
ROM+SR & $u_x$       & 17   & $0.45 \pm 0.16 \%$ &   $0.74 \pm 0.57 \%$     &   1.17 $\times 10^{-4}$    &     \multirow{3}{*}{1.42}          \\
\cmidrule(lr){1-6}
ROM+SR &  $u_y$        & 21 & $0.83 \pm 0.31 \%$ &   $1.62 \pm 1.52\%$     & 8.53 $\times 10^{-5}$  &          \\
\cmidrule(lr){1-6}
ROM+SR &  $p$        & 8   & $1.37 \pm 0.50\%$  &  $2.73\pm 2.22\%$     & 1.28 $\times 10^{-4}$  &      \\
\bottomrule
\end{tabular}
\end{adjustbox}
\caption{Error estimation and computational time of the FOM and proposed surrogate model based on ROM with geometry-informed snapshots. The relative $L^2$ error of POD and the entire surrogate model and the computational time were evaluated based on the test dataset with 50 samples and $99.9\%$ cumulative energy. The computational time of prediction consists of two parts: the average computational time of surface registrations (SR) and non-intrusive reduced order model (ROM) evaluations. Note that the FOM here denotes performing a simulation with FOM on its original domain. All the predictions here were performed on Intel Core i7-7500U CPU@2.70GHz.}
\label{tab:speedup_case2}
\end{table}

{\renewcommand{\arraystretch}{1.2}
\begin{table}[th!]
\centering
\begin{adjustbox}{width=1\textwidth}
\begin{tabular}{ccc|cccc|c}
\toprule
\multicolumn{3}{c|}{Registration} 
& \multicolumn{4}{c|}{ROM}  
& Other  \\ \hline
\multicolumn{1}{c|}{\begin{tabular}[c]{@{}c@{}}surface \\ registration\end{tabular}} 
& \multicolumn{1}{c|}{\begin{tabular}[c]{@{}c@{}}geometric parameters \\ compression (SVD)\end{tabular}} 
& \begin{tabular}[c]{@{}c@{}}diffeomorphism \\ appproximation (RBFI)\end{tabular} 
& \multicolumn{1}{c|}{properties} 
& \multicolumn{1}{c|}{\begin{tabular}[c]{@{}c@{}}reduced basis \\ construction (SVD)\end{tabular}} 
& \multicolumn{1}{c|}{\begin{tabular}[c]{@{}c@{}}reduced coefficient \\ mapping (RBFI)\end{tabular}} 
& \begin{tabular}[c]{@{}c@{}}Snapshot \\ generation\end{tabular} 
& Data processing  \\ \hline

\multicolumn{1}{c|}{\multirow{3}{*}{$1.42N_s $}}  
& \multicolumn{1}{c|}{\multirow{3}{*}{0.16}}   
& \multirow{3}{*}{0.072}                                                               
& \multicolumn{1}{c|}{$u_x$}      
& \multicolumn{1}{c|}{$0.74 $}                                                        
& \multicolumn{1}{c|}{$0.024 $}                                                         
& \multirow{3}{*}{41.2$N_s$}                                              
& \multirow{3}{*}{262.5  } \\ \cline{4-6}
\multicolumn{1}{c|}{}                                                                
& \multicolumn{1}{c|}{}                                                                                  &                                                                                
& \multicolumn{1}{c|}{$u_y$}      
& \multicolumn{1}{c|}{$0.75 $}                                                        
& \multicolumn{1}{c|}{$0.021 $}                                                         
&                                                               
&                  \\ \cline{4-6}
\multicolumn{1}{c|}{}                                                               
& \multicolumn{1}{c|}{}                                                                                  &                                                                                 
& \multicolumn{1}{c|}{$p$}        
& \multicolumn{1}{c|}{$0.69 $}                                                        
& \multicolumn{1}{c|}{$0.041$}                                                         
&                                                                
&                   \\ \bottomrule
\end{tabular}
\end{adjustbox}
\caption{Computational time (s) of offline components of the proposed method for the bifurcation case. $N_s =400$ denotes the number of data used for training. All the computations here were performed on Intel Core i7-7500U CPU@2.70GHz, except that the snapshot generation was performed on a shared AMD Rome 7H12 node with 16-core on Dutch supercomputer Snellius.}
\label{tab:offline2}
\end{table}}
    
The surface registration was performed to compute the geometric parameters and the corresponding mapping between the reference shape and other shapes. The study of kernel width shown in Figure~\ref{fig:BifurPOD}(a) indicates that the average residual of the registration is lowest when it is set to be 1.6 and the average residual of the registration is $3.6 \times 10^{-6}$.  There are 580 nodes on the boundary of the reference shapes which leads to 1160 geometric parameters in the two-dimensional case. Owing to the POD, the dimension of the geometric parameters was reduced to $8$, which matches the dimensions of parameters used for shape generation. It is obvious that both $L^2$ error and the truncated energy visualised in Figure~\ref{fig:BifurPOD}(b) had a sharp drop with the first eight bases and became flattened after. 
The accuracy of the coordinate mapping is again measured by the accuracy of the result solved on the reference domain. The average relative $L^2$ errors of velocities in x and y directions, and pressure of 400 training samples are $0.033\%$, $0.069\%$, and $0.026\%$, respectively.

The bases of the ROM were subsequently constructed. The velocities in the x and y directions require $17$ and $21$ reduced bases to capture 99.9\% of the total variance of the data, while pressure needs $8$ reduced bases. Figure~\ref{fig:BifurPOD}(c) and (d) demonstrate the cumulative energy and its corresponding error introduced by POD against the number of bases. The predictions of the ROM for one of the test cases are shown in Figure~\ref{fig:chap4bifur_pred}. The average $L^2$ errors of velocities in the x and y directions and pressure over 50 validation cases are $0.70\%$, $1.53\%$, and $2.65\%$, respectively. The error estimations and computational time of the test dataset are presented in Table~\ref{tab:speedup_case2}. Similar to the previous case, the main computational cost of the surrogate model comes from surface registration. The offline computational time is evaluated and presented in Table~\ref{tab:offline2}. 

As shown in Table~\ref{tab:speedup_case1} and \ref{tab:speedup_case2}, the major discrepancy of the surrogate model originated from the POD. Nonetheless, a further increase of the reduced bases to cover more cumulative energy of the snapshots doesn't necessarily improve the performance of the surrogate model. Figure~\ref{fig:PODincrease}  illustrates the variation in the relative $L^2$ error of prediction with the increase of cumulative energies in POD. In the case of stenosis, 
the surrogate model discrepancy decreased slightly with the increase of cumulative energy from $99.9\%$ to $99.99\%$ and barely improved when increased from $99.99\%$ to $99.999\%$. On the other hand, the surrogate model discrepancy of the bifurcation case stopped decreasing after $99.9\%$. This is mainly because  the high-frequency bases, characterized by small singular values, tend to be challenging to interpolate (e.g. the last coefficient shown in Figure~\ref{fig:chap4RC}).

\section{Discussion}\label{sec:discussion}
The error of the surrogate model mainly stems from three aspects: registration, decomposition, and interpolation. The diffeomorphisms, constructed via registration, enable the systems to be solved on a reference domain. The residual in the registration hence introduced an error into diffeomorphism and subsequently propagated to snapshots. The error in the two case studies was reasonably small as the registration aligned well enough, therefore could be neglected. However, achieving an accurate diffeomorphism through registration between two general shapes is a non-trivial task. Here we emphasise the assumption that the geometries can be achieved by small perturbation from the reference shape.

The second part of the error arises from POD where a number of reduced bases was selected to represent the original data. The criteria was that the cumulative energy captured by $N_l$ of reduced bases reached 99.9\%. The singular values of the SVD decayed rapidly as shown in Figure~\ref{fig:chap4POD}(c) and Figure~\ref{fig:BifurPOD}(c). The error introduced by POD is the majority of the overall discrepancy, as shown in Table~\ref{tab:speedup_case1}, Table~\ref{tab:speedup_case2}. To further reduce the discrepancy introduced by the Proper Orthogonal Decomposition, there are two main approaches. First, expand the number of reduced bases to capture more data variance. However, since the singular values of the high-frequency bases decay slowly, significantly more bases will be involved and the latent functions of the coefficients of those high-frequency bases are typically less smooth and challenging to interpolate. Therefore, it will not necessarily improve the prediction of the surrogate model, as shown in Figure~\ref{fig:PODincrease}(b). On the other hand, the discrepancy of POD is also constrained by the number of snapshots available for training. These snapshots represent discrete approximations of the solution manifolds. Further improvement can be achieved by either including more snapshots or choosing the snapshots selectively with certain criteria, such as the greedy method \cite{Quarteroni2011} or active learning \cite{KAST2020112947}. Note that unlike parametric problems only involving physical parameters, where a new snapshot simply means another evaluation of the model with particular new parameters, geometrical problems may also require you to generate a domain with a corresponding new shape artificially.

The third part of the error originates from the non-intrusive prediction with RBF interpolation. The RBF interpolation can be replaced by other interpolation methods such as polynomial interpolation \cite{Gasca2000} or a neural network \cite{goodfellow2016deep}, etc. The error of this part should be first evaluated and tuned against a validation dataset before employing the trained interpolator for prediction. Similar to POD, the error of the RBF interpolator also significantly depends on training data. Note that the boundary condition is persevered by the construction of the ROM, therefore the velocity error on the boundary is intrinsically small. However relatively large errors may occur on the boundary of pressure prediction since no boundary condition of pressure is enforced.

The computational time of the surrogate model for the examples was demonstrated in Table~\ref{tab:speedup_case1} and \ref{tab:speedup_case2}. A speedup of around 3 and 10 (not taking into account the time required for training) for online prediction was achieved in prediction by applying non-intrusive ROM and surface registration for each case respectively. The computational cost in prediction mainly came from surface registration while the non-intrusive ROM barely takes any time as the operation involved is weighted summations. This also enables the possibility of real-time performance in clinical practice owing to the separation of online and offline processes since the main computational effort was distributed to the offline part. However, the speedup shown here is only based on two-dimensional synthetic examples. Generally, the two-dimensional problem itself would be computationally cheap enough for a direct simulation rather than spending a huge amount of time on offline preparation. A three-dimensional problem with a complex domain e.g. image-based patient-specific simulation is much more computationally demanding and therefore closer to the purpose of the method. The corresponding time for prediction also increases. The computational time of surface registration increases almost linearly in log-log scale with the number of boundary vertices if a proper parallelisation is implemented on CPU/GPU as shown in \cite{deformatrica}. Therefore, a further investigation on the speedup for a three-dimensional patient-specific dataset would  be required and we aim to implement three-dimensional examples in our future work. Nevertheless, the examples prove the concept of the proposed method and demonstrate its accuracy and efficiency.

The surrogate model proposed in this work is in the context of biomedical science and engineering, mainly aiming at hemodynamics. We assume that with state-of-art imaging and segmentation techniques, a large amount of geometric data of arteries are available. Instead of handling each geometric data to perform a simulation individually, the geometric dataset of a kind (similar but distinct) can be leveraged for further prediction based on ROM. The proposed method can be also applied to other physiological flow simulations, e.g. respiratory flow or cerebrospinal flow as long as the surface registration can be performed between the different shapes of the domains and construct corresponding diffeomorphisms.  Likewise, such a method of course is not limited to biomedical applications. For example in environmental science, fluvial bank erosion simulation always needs to solve the fluid dynamics problem with changing domains over time. Other fields such as design optimization especially shape optimization in automotive and aerospace engineering also require constant evaluations of fluid dynamics with similar shapes of domains.

In this work, we limit the scope of the problem to the physiological fluid system with moderate Reynolds’s number and steady situation. In practice, we will face more complex situations such as problems with time-dependency (e.g. pulsatile flow) or with a wide range of Reynolds numbers depending on the sections of the arteries (e.g. from $Re\approx1$ in arterioles to $Re\approx4000$ in the aorta). For time-dependent cases, the construction of the snapshot matrix would also include the changes of variable fields due to time evolution and an extra dimension for temporal coordinates should be added to the input space for the non-intrusive ROM interpolator. Different Reynolds numbers will lead to completely different behaviours of fluid. For instance, at Reynolds number around 2000, the laminar flow starts to transition into turbulent flow and becomes fully turbulent in the aorta \cite{Mahalingam2016}. We leave the study of those more complex situations to our future work.
% Note that this work is presented in the context of blood flow simulations with steady-state. In practice, more complex situations with time dependency (e.g. pulsatile flow) will be encountered. The transient simulations of course capture more details and characteristics of the physiological flow, such as the problems involving mass transport \cite{LIU20111123} or stagnation \cite{Kar2005,Corbett2010}. However, steady-state simulations with time-averaged velocity in many scenarios have provided enough information on hemodynamics to facilitate investigation and understanding of physiological and pathological problems \cite{Geers2010,MULLER20122539,HSIAO2012128}. A further discussion can be found in Section~\ref{sec:discussion}.

\section{Conclusion}\label{sec:conclusion}
In this work, a surrogate model based on surface registration and non-intrusive reduced-order modelling is proposed. The surface registration with currents provides the diffeomorphism between two coordinate systems and parameterises the shape without point-to-point correspondence. With the diffeomorphisms, all the evaluations of FOM are subsequently performed on a reference domain, embedding the geometry information in the snapshots and guaranteeing the spatial compatibility of snapshots. The non-intrusive reduced-order model is subsequently constructed using POD and the RBF interpolator is trained for predicting the reduced coefficients of ROM based on reduced geometric parameters of the shape. Two examples of blood flow simulations based on stenosis vessels and bifurcation vessels are presented and discussed.

\section*{Funding}
\noindent This work has received funding from the European Union Horizon 2020 research and innovation programme under grant agreements \#800925 (VECMA project). This work was sponsored by NWO Exacte Wetenschappen (Physical Sciences) for the use of supercomputer facilities, with financial support from the Nederlandse Organisatie voor Wetenschappelijk Onderzoek (Netherlands Organization for Science Research, NWO).

% \section*{Authors' contributions}
% \noindent D. Ye has designed and developed the surrogate model for blood flow simulation under supervision of A.G. Hoekstra and V. Krzhizhanovskaya. D. Ye has drafted manuscript. All authors have revised the paper.
\section*{Data availability}
All the data and source codes to reproduce the results in this study are available on GitHub at \url{https://github.com/DongweiYe/ROM-with-geometry-informed-snapshots}

\section*{Declaration of competing interest}
\noindent The authors declare that they have no known competing financial interests or personal relationships that could
have appeared to influence the work reported in this paper.

\bibliographystyle{vancouver}
% \bibliographystyle{elsarticle-num}
% \bibliography{Reference.bib}

\end{document}